\documentclass[aps,reprint]{revtex4-2}
\usepackage{blindtext}
\usepackage{amsmath}
\usepackage{amssymb}
\usepackage{graphicx}
\usepackage{textcomp}
\newcommand{\Vin}{\text{\textcent} } 

\def\a{\alpha}
\def\b{\beta}
\def\r{\rho}

\def\l{\lambda}

\def\t{\tau}

\def\a{\alpha}
\def\b{\beta}

\def\G{\Gamma}

\def\rd{{\rm d}}

\def\cD{{\mathcal{D}}}
\def\cP{{\mathcal{P}}}
\def\cK{{\mathcal{K}}}

\def\cN{{\mathcal{N}}}

\def\bcP{{\bar{\cP}}}
\def\bnabla{{\bar{\nabla}}}

\usepackage{appendix} 

\def\bbE{\mathbb{E}}

\def\tbbE{\tilde{\mathbb{E}}}

\begin{document}

        \title{Diffeomorphism Radiative Degrees of Freedom of Thomas-Whitehead Gravity}
        \author{Owen Fiedorowicz}
        \email{owen-fiedorowicz@uiowa.edu}
        \author{Tyler C. Grover}
        \email{tyler-grover@uiowa.edu}
        \author{Vincent G. J. Rodgers}
        \email{vincent-rodgers@uiowa.edu}
        \affiliation{University of Iowa}
        
        \author{Hazal D. Zenger}
        \email{zengerha@grinnell.edu}
        \affiliation{Grinnell College}
        
        \begin{abstract}
                The geometric action of the semi-direct product of the Kac-Moody and Virasoro  algebras contains the WZW action equipped with a background  vector potential $A$ associated to a coadjoint element of the Kac-Moody algebra as well as the  2D gravitational  Polyakov action and an accompanying background field, $\cD$, called the diffeomorphism field.  Just as the coadjoint element, $A$, is related to a gauge fixed Yang-Mills vector potential $A_a$, the diffeomorphism field, $\cD$, is related to a component, $\cD_{a b}$  of the projectively invariant connection called the Thomas Operator. The Yang-Mills action provides dynamics for the vector potential $A_a,$ while the Thomas-Whitehead (TW) gravitational action, provides dynamics to $\cD_{ab}$. The TW action embeds the projectively invariant connection into a gravitational  theory that contains general relativity \cite{GS,DPC,RodgersReview2022}. In this work, the diffeomorphism field $\cD_{a b}$ is examined in Minkowski space where salient features of this field can be explored.  In particular, we study the radiative degrees of freedom of this field while in a Minkowski space background. We show that it can be  decomposed into irreducible representations, corresponding to  tensor, vector, and scalar radiating solutions. Furthermore we examine geodesic  deviation in the context of  TW gravity about a Minkowski space background.  We do this both at zeroth and first order  in metric fluctuations  $h_{ab}$. We discuss that response of a gravitational wave antennae to the geodesics deviations. 
        \end{abstract}
        
        \maketitle

        \section{Introduction}

        The motivation of this work is to understand the radiative degrees of freedom of the diffeomorphism field and its effect on geodesic deviation. Contributions to geodesic deviation from either the metric or $\Pi$ are minimal. We will start with the TW action and its field equations. The diffeomorphism field equations will then be examined in a Minkowski space background. A cosmological constant is incorporated for Minkowski space to be an exact solution when $\mathcal{P}_{ab} = \mu g_{ab}$, where $\mathcal{P}_{ab } $ (see Eq. \ref{Pdefinition}) is the tensor extension of the connection component $\cD_{ab}.$  This will relate the bare cosmological constant $\Lambda$ to  the metric trace of $\mathcal{P}_{ab}.$ The diffeomorphism field will then be perturbed about this expectation value, with its dynamics given by the remaining homogeneous diffeomorphism field equation. The perturbation is then decomposed into trace and traceless parts. The traceless part being further decomposed into a transverse-traceless and longitudinal-traceless parts, and the longitudinal-traceless part being even further decomposed into solenoidal and scalar components.  The diffeomorphism field fluctuation then has a full decomposition in terms of transverse-traceless, solenoidal, and trace components. The TW action in  Minkowski space along with the  Levi-Civita connection gives  us a collection of  massive free field theories in the Minkowski space background. To first order, we examine the fiducial geodesics about the  $\mathcal{P}_{ab} = \mu g_{ab}$ solution. Geodesic deviation about these fiducial geodesics is considered by having $\tilde{\Gamma}$, the projective connection,  sourced by the massive field fluctuations. The resulting deviation is trivial in a pure Minkowski space with a Levi-Civita connection. We then consider the first order metric fluctuation $h_{ab}$ sourced from the massive diffeomorphism field fluctuations. The geodesic deviation from the first order metric fluctuation $h_{ab}$ is then computed numerically, and antenna patterns for a LIGO-like interferometer for all of the polarizations considered are provided.
        
\subsection{TW Theory}

Thomas-Whitehead (TW) gravity is a gauge theory of gravity with respect to the 
equivalence of paths under projective transformations of affine connection.  If one thinks of a $d$ dimensional  Riemannian manifold in terms of its locus of geodesics, in other words its \emph{development}, one finds that there is an ambiguity.  This follows since geodesics, that satisfy  $$\xi^\a \nabla_a \xi^b = f(x) \xi^b$$ and geodetics that satisfy $$\eta^\a \nabla_a \eta^b =0,$$are related through reparameterizations of the geodesic paths as well rescaling of the vector fields.  Furthermore, another connection, $\hat \nabla_a$, could also render $\xi^a$ a geodetic if there is a suitable \emph{projective transformation} of the connection symbols, \[\G^{a}_{\,\,b c} = {\hat \G}^{a}_{\,\,b c}+ \delta^a_b v_c + \delta^a_c v_b,\] for a particular $v_a$.  Then $\xi^\a \hat\nabla_a \xi^b =0, $ leaving $\xi^a$ a geodetic.   Tracey Yerkes Thomas  \cite{Thomas1,Thomas2} developed a geometry for the equivalent classes of connections, $[\G^{a}_{\,\,b c}], $  that were projectively related by introducing  $\Pi^a{}_{bc} $, the fundamental projective invariant. This symmetry is then realized as a volume bundle over the original manifold.  The new $d+1$ dimensional structure is now called the \emph{Thomas cone} and denoted by  $\cN$. $\cN$ is equipped with its own projectively invariant connection, $\tilde \G^{\a}_{\,\,\b \r} (\tilde \nabla_\a),$ and  is covariant with respect to transformations that are in one-to-one correspondence with the $d$ dimensional general coordinate transformation. The Greek indices now span $d+1$  coordinates.  \  Thomas achieved covariance   on  $\cN$ by introducing what he called the projectively invariant Ricci symbol ${\cal R}_{a b}$ that is the usual Ricci tensor built from $\Pi^a{}_{bc} $ instead of the connection $\G^{a}_{\,\,b c}$.  The non-tensorial transformation laws of ${\cal R}_{a b}$ and $\Pi^a{}_{bc} $ conspire to guarantee that $\tilde \G^{\a}_{\,\,\b \r} (\tilde \nabla_\a)$ is covariant with respect to the transformations of the $d+1$ coordinates discussed in the Appendix (\ref{TCTransformations}). This work was generalized in Ref.\cite{DPC} by replacing ${\cal R}_{a b}$ with what is known as the diffeomorphism field, $\cD_{ab}$.  The diffeomorphism field transforms just as ${\cal R}_{a b}$ and  has its origins in 2D quantum gravity due to its association with the coadjoint elements of the Virasoro algebra.  The one dimensional reduction of $\cD_{ab}$ shows that it transforms as a coadjoint element of the Virasoro algebra.  Such a reduction to one dimension is not possible for  ${\cal R}_{a b}$. This association is exactly parallel to the relationship between the  coadjoint elements of   Kac-Moody algebras and  gauge potential $A_\mu$ in higher dimensions \cite{CO}.  While the Kac-Moody algebra enjoys higher dimensional realizations as Yang-Mills theories, an equivalent generalization for the diffeomorphism field was sought. The diffeomorphism field is now realized as connection component of a d+1 dimensional manifold $\cN$.  TW gravity generalizes the work of Thomas by incorporating $\cD_{ab}$ as an independent connection component along with  $\Pi^a{}_{bc}$ while maintaining covariance of the covariant derivative operator.  TW gravity is built from curvature invariance, $\mathcal{K}^{\alpha}{}_{\beta\sigma\rho},$ derived from the covariant derivative operator $(\tilde \nabla_\a).$ The curvature squared contributions in the action give $\cD_{ab}$ dynamics akin to curvature squared theories for Yang-Mills. One can show that TW gravity reduces to Einstein's General Relativity when $\cD_{ab}=0$ and $\Pi^a{}_{bc} $ is the equivalence class for the Levi-Civita connection of the metric.  A review of TW gravity can be found in \cite{GS}. For convenience, a summary of definitions and identities is provided in the appendix. More recently \cite{EF}, TW gravity has been formulated in terms of fields that are both tensorial and projectively invariant. The TW action may be written as,
\begin{align} 
                \label{e:TWActionExpanded}
                S = &  -\frac{1}{2 \kappa_0} \int d^\rd x\sqrt{|g|} \left(\mathcal{K}+2\Lambda_0\right) \cr & + J_0 c \lambda_0^2 \int d^\rd x\sqrt{|g|} \left( K_{bcd} K^{bcd} \right)   \cr &  - J_0 c \int \sqrt{|g|}d^\rd x\left( \mathcal{K}_{abcd} \mathcal{K}^{abcd} - 4 \mathcal{K}_{bd} \mathcal{K}^{bd} + \mathcal{K}^2 \right), 
        \end{align}
         with coefficient $\kappa_0$ $J_0$, and $\lambda_0$.  This admits the field equations for $\cD_{ab} $ viz,
\begin{align}
\lambda_0^2\bar{\nabla}_{d}K^{(ij)d}& - \left(d-1\right)\mathcal{K}g^{ij} \cr  &+2\left(2d-3\right)\mathcal{K}^{ij}-\tfrac{d-1}{4\kappa_0J_0c}g^{ij} = 0.
        \end{align}
Similarly the field equations for $\Pi^a{}_{bc }$ are,
        
        \begin{align}\label{e:EOMC}
                \tbbE_c{}^{ab} \equiv & \bbE_c{}^{ab} - \tfrac{1}{\rd+1} \delta_c{}^{(a} \bbE_m{}^{b)m} = 0 \\
                \mathbb{E}_c{}^{ab} =& \bnabla_c\left[ g^{ab} \left(\cK + \tfrac{1}{4 J_0 \kappa_0}\right) - 4 \cK^{ab}\right] - \bnabla_d \left(g_{cm} \cK^{m(ab)d}\right) \cr & - \lambda_0^2 \bcP_{dc} K^{(ab)d}, \end{align}
while the field equations for $g_{ab}$ yield,
 \begin{align}\label{e:cHmn}
                & -\mathcal{K}_{mn} -  \kappa_0  g_{mn} \left[\tfrac{2 J_0 \l_0^2}{(\rd+1)} \bar{\nabla}_a (\cK^{a}{}_{bcd}K^{bcd})+\mathcal{L} \right] \cr
&+ 2 \kappa_0 J_0 \left( \lambda_0^2 \left(K_{mcd}K_n{}^{cd} + 2 K_{bcm}K^{bc}{}_n \right)\right. \cr
&+\cK_{mbcd}\cK_n{}^{bcd} -\cK^a{}_{mcd}\cK_{an}{}^{cd} - 2 \cK^{abc}{}_m \cK_{abcn}\cr  & + \left. 8 \,\cK_{mb}\cK^b{}_n - 2 \cK \cK_{mn}\right)  = ~0,
        \end{align}
where $\mathcal{L}$ is the TW Lagrangian, 
\begin{align}\label{e:cHmn1}
                \mathcal{L} =& - \tfrac{1}{2\kappa_0} \left(\mathcal{K}+2\Lambda_0\right)+ J_0c\lambda_0^2 K_{bcd}K^{bcd}\cr
                & - J_0c(\mathcal{K}_{abcd} \mathcal{K}^{abcd} - 4 \mathcal{K}_{bd} \mathcal{K}^{bd} + \mathcal{K}^2).
        \end{align}
    \subsection{(A)dS Solution}
        
        In \cite{DE,GS} it was shown that the TW Field equations admit a one parameter family of solutions which we refer to as the (Anti) de Sitter solutions. In these solutions, the metric is for an Einstein manifold, the projective Schouten tensor is proportional to the metric by the parameter $\mu$, and the connection is Levi-Civita. The field equations then relate the parameter $\mu$ to the Ricci scalar while simultaneously determining the Cosmological constant $\Lambda_0$. More succinctly:
\begin{align}
                \begin{split}
                        C^a{}_{bc} = 0,~~ g_{ab} &= \tfrac{d-2}{2\Lambda}R_{ab},~~      \mathcal{P}_{ab} = \mu g_{ab}\\
                        \Lambda_0 = \tfrac{3}{8J_0\kappa_0c},~~ &\Lambda = \tfrac{3}{8J_0\kappa_0c} + 3\mu
                \end{split}
        \end{align}
        
        \subsubsection{Minkowski Space}
        
        Minkowski space is then a member of this solution space for parameter $\mu = \tfrac{-1}{8J_0\kappa_0c}.$
\begin{align}
                \begin{split}
                        C^a{}_{bc} = 0,~~ g_{ab} &= \eta_{ab}.~~ \mathcal{P}_{ab} = \tfrac{-1}{8J_0\kappa_0c} \eta_{ab}\\
                        \Lambda_0 = \tfrac{3}{8J_0\kappa_0c},~~ &\Lambda = 0
                \end{split}
        \end{align}
        
        \subsection{Diffeomorphism Field Equations}
        
        In subsequent sections, we will consider the TW action in a Minkowski space background with a compatible connection. Then only the dynamics of the diffeomorphism field need to be considered. For now, we'll expand the diffeomorphism field equation into an inhomogeneous equation where the homogeneous part considers the metric to be a fluctuation about Minkowski space. This will provided the machinery for calculating higher order corrections to the diffeomorphism field from the back reaction of the metric and the so-called Palatini tensor which allows the connection to deviate from Levi-Civita . The diffeomorphism field equations can be cast into the form of inhomogeneous linear differential equations with inhomogeneous source $S_{xy}.$
       \begin{align}
                \begin{split}
                        &\lambda_0^2\left(\eta^{di}\partial_d\partial_x\mathcal{P}_{yi} + \eta^{di}\partial_d\partial_y\mathcal{P}_{ix} - 2\eta^{zd}\partial_d\partial_z\mathcal{P}_{xy}\right)\\
                        & + 2\left(2d-3\right)\left(d-1\right)\mathcal{P}_{xy} - \left(d-1\right)^2\eta^{ab}\eta_{xy}\mathcal{P}_{ab}\\
                        & = S_{xy}
                \end{split}
        \end{align}
 The components of $S_{xy}$ can collected in terms of contributions from terms that have no $\mathcal{P}$ dependence, connection terms, inhomogeneous contributions from each of the curvature tensors. 
\begin{align}
                \begin{split}
                        S_{xy} = &(S_1)_{xy} + \left(d-1\right)(S_3)_{xy}\\
                        & - \lambda_0^2g_{cx}g_{by}\partial_dS^{bcd}_2 + (S_4)_{xy}
                \end{split}
        \end{align}
$S_1$ has the inhomogeneous contributions from the metric term, the projective Ricci tensor, and the connection terms from $\nabla_{d}K^{(ij)d}.$
\begin{align}
                \begin{split}
                        (S_1)_{xy} = &\frac{1}{4\kappa_0J_0c}\left(d-1\right)g_{xy} - \lambda_0^2g_{cx}g_{by}\bar{\Gamma}^{b}{}_{di}K^{(ic)d}\\
                        & - \lambda_0^2g_{cx}g_{by}\bar{\Gamma}^{c}{}_{di}K^{(bi)d} - \lambda_0^2g_{cx}g_{by}\bar{\Gamma}^{d}{}_{di}K^{(bc)i}\\
                        & - 2\left(2d-3\right)\hat{R}_{xy}
                \end{split}
        \end{align}
$S_2$ has the inhomogeneous contributions from $K^{(ij)d}$ after stripping the connection terms from $\nabla_{d}K^{(ij)d}.$
\begin{align}
                \begin{split}
                        S^{pqc}_2 = &\epsilon h^{yq}\partial_y\mathcal{P}^{pc} + \epsilon h^{yp}\partial_y\mathcal{P}^{cq} - 2\epsilon h^{zc}\partial_z\mathcal{P}^{pq}\\
                        & - g^{x(p|}g^{y|q)}g^{zc}\Gamma^i{}_{yz}\mathcal{P}_{ix} - g^{x(p|}g^{y|q)}g^{zc}\Gamma^i{}_{yx}\mathcal{P}_{zi}\\
                        & + g^{x(p|}g^{y|q)}g^{zc}\Gamma^i{}_{zy}\mathcal{P}_{ix}
                         + g^{x(p|}g^{y|q)}g^{zc}\Gamma^i{}_{zx}\mathcal{P}_{yi}\\
                \end{split}
        \end{align}
$S_3$ has the inhomogeneous contributions from the projective Ricci Scalar, $K.$
        
        \begin{align}
                \begin{split}
                        &(S_{3})_{xy} = \hat{R}\left(\eta_{xy} + \epsilon h_{xy}\right)\\
                        & + \left(d-1\right)\left(\epsilon \bar{h}^{ab}\eta_{xy} + \epsilon\eta^{ab}h_{xy} + \epsilon^2h_{xy}\bar{h}^{ab} + \mathcal{O}(\epsilon^3)\right)P_{ab}
                \end{split}
        \end{align}
        
        $S_4$ has the inhomogeneous contributions from the metric $g_{ab}$ in $\partial_{d}K^{(ij)d}.$
        
        \begin{align}
                &(S_4)_{xy} = \cr &- \lambda_0^2\left(\eta_{cx}h_{by} + \eta_{by}h_{cx} + h_{cx}h_{by}\right)\cr
                &\times\partial_d\left(\eta^{yc}\partial_y\mathcal{P}^{bd} + \eta^{yb}\partial_y\mathcal{P}^{dc} - 2\eta^{zd}\partial_z\mathcal{P}^{bc}\right)
        \end{align}
The diffeomorphism field solution is then the homogeneous solution with a particular solution from evaluating the Green function on the source. The homogeneous solution $\mathcal{P}_{H}$ solves
        
        \begin{align}
                \begin{split}
                        &\lambda_0^2\left(\eta^{dk}\partial_d\partial_x(\mathcal{P}_{H})_{ky} + \eta^{dk}\partial_d\partial_y(\mathcal{P}_{H})_{xk} - 2\eta^{dk}\partial_d\partial_{k}(\mathcal{P}_{H})_{xy}\right)\\
                        &- \left(d - 1\right)^2\eta_{xy}\eta^{ab}(\mathcal{P}_{H})_{ab} + 2\left(2d - 3\right)\left(d-1\right)(\mathcal{P}_{H})_{xy} = 0.\\
                \end{split}
        \end{align}
The full solution incorporating the particular solution is then
\begin{align}
                \begin{split}
                        \mathcal{P}_{pq} &= (\mathcal{P}_{H})_{pq} + \int d^4x'\sqrt{|g|}G_{pq}{}^{uv}(x,x')S_{uv}(x'),
                \end{split}
        \end{align}
where the Green function satisfies
\begin{align}
                \begin{split}
                        &\lambda_0^2\left(\eta^{dk}\partial_d\partial_xG_{ky}{}^{uv} + \eta^{dk}\partial_d\partial_yG_{xk}{}^{uv} - 2\eta^{dk}\partial_d\partial_{k}G_{xy}{}^{uv}\right)\\
                        &- \left(d - 1\right)^2\eta_{xy}\eta^{ab}G_{ab}{}^{uv} + 2\left(2d - 3\right)\left(d-1\right)G_{xy}{}^{uv}\\
                        & = \delta^{u}{}_{(x}\delta^{v}{}_{y)}\delta^{(4)}(x,x').\\
                \end{split}
        \end{align}
Now, we take the Fourier transform to specify the propagator $\tilde{G}$ associated with $G.$
\begin{align}
                \begin{split}
                        &\lambda_0^2\left(\eta^{dk}k_dk_x\tilde{G}_{ky}{}^{uv} + \eta^{dk}k_dk_y\tilde{G}_{xk}{}^{uv} - 2\eta^{dk}k_dk_{k}\tilde{G}_{xy}{}^{uv}\right)\\
                        &- \left(d - 1\right)^2\eta_{xy}\eta^{ab}\tilde{G}_{ab}{}^{uv} + 2\left(2d - 3\right)\left(d-1\right)\tilde{G}_{xy}{}^{uv}\\
                        & = \delta^{u}{}_{(x}\delta^{v}{}_{y)}\\
                \end{split}
        \end{align}
The algebra constraining $\tilde{G}$ is formed from $\eta,$ k, and $\delta.$ This informs the shape of $\tilde{G}$ so that
\begin{align}
                \begin{split}
                        \tilde{G}_{ij}{}^{uv} = &A\eta_{ij}\eta^{uv}+ B\eta_{ij}k^{u}k^{v}+ C\eta^{uv}k_{i}k_{j}+ D\delta^{u}{}_{(i}\delta^{v}{}_{j)}\\
                        &+ E\delta^{(u}{}_{(i}k_{j)}k^{v)}+ Fk_ik_jk^uk^v.
                \end{split}
        \end{align}
 Enforcing this ansatz on the propagator equation determines the coefficients. For arbitrary dimension d, the coefficients are related by
        
        \begin{align}
                &D = \tfrac{1}{2\alpha_1}\\
                &E = \tfrac{-\lambda_0^2}{\alpha_2\alpha_1}\\
                &C = \tfrac{-\left(d-1\right)\lambda_0^2}{2\left(2d-3\right)\alpha_3\alpha_1}\\
                &A  = \tfrac{\left(d-1\right)^2}{2\alpha_3\alpha_1}\\
                &B = \tfrac{1}{\alpha_3}E\left(d-1\right)^2\left(1 - \tfrac{\lambda_0^2k^2}{2\left(2d-3\right)\left(d-1\right)}\right)\\
                &F = \tfrac{-2B\lambda_0^2 - E\lambda_0^2}{2\left(2d-3\right)\left(d-1\right)}
        \end{align}
 where
\begin{align}
                \alpha_1 &= \left(2d-3\right)\left(d-1\right) - \lambda_0^2k^2 \\
                \alpha_2 &= -\lambda_0^2k^2 + 2\left(2d-3\right)\left(d-1\right)\\
                \alpha_3 &= 2\left(2d-3\right)\left(d-1\right)  - \left(d-1\right)^2d\cr
                & + \tfrac{\left(\left(d-1\right) - 2\left(2d-3\right)\right)\lambda_0^2k^2}{\left(2d-3\right)}
        \end{align}
These terms in the propagator contain the  poles
\begin{align}
                &k_1 = \pm i\tfrac{\sqrt{\left(2d-3\right)\left(d-1\right)}}{\lambda_0};~ k_2 = \pm i\tfrac{\sqrt{2\left(2d-3\right)\left(d-1\right)}}{\lambda_0}\\
                &k_3 = \pm\tfrac{i}{\lambda_0}\sqrt{\tfrac{\left(2d-3\right)\left(d\left(d-1\right)^2-2\left(2d-3\right)\left(d-1\right)\right)}{\left(d-1\right)-2\left(2d-3\right)}}.
        \end{align}
For 4d spacetime, the coefficients and poles become
\begin{align}
                A &= \tfrac{45}{7\lambda_0^4k^4+75\lambda_0^2k^2-450}\\
                B &= \tfrac{3\lambda_0^2}{7\lambda_0^4k^4+75\lambda_0^2k^2-450}\\
                C &= \tfrac{3\lambda_0^2}{7\lambda_0^4k^4+75\lambda_0^2k^2-450}\\
                D &= \tfrac{1}{\lambda_0^2k^2+15}\\
                E &= \tfrac{2\lambda_0^2}{7\lambda_0^4k^4+45\lambda_0^2k^2+450}\\
                F &= \tfrac{2\lambda_0^2\left(\lambda_0^2k^2+6\right)}{3\left(7\lambda_0^2-30\right)\left(\lambda_0^2+30\right)\left(\lambda_0^2+15\right)}
        \end{align}
\begin{align}
                &k_1 = \pm i\tfrac{\sqrt{15}}{\lambda_0};~ k_2 = \pm i\tfrac{\sqrt{30}}{\lambda_0};\\
                &k_3 = \pm\tfrac{1}{\lambda_0}\sqrt{\frac{30}{7}}.
        \end{align}
        
        \subsection{$\mathcal{P}_{ab}$ Perturbation}
        
        In the interest of understanding the simplest novel aspects of TW gravity on gravitation, we entertain an approximation where we consider fluctuations $\hat{\mathcal{P}}_{ab}$ in the diffeomorphism field about the Minkowski space solution in a fixed Minkowski space background metric with Levi-Civita connection. Cartesian coordinates are chosen for convenience. With this we may write,
\begin{align}
                \mathcal{P}_{ab} = \tfrac{-1}{8J_0\kappa_0c} \eta_{ab} + \hat{\mathcal{P}}_{ab}.
        \end{align}
        These fluctuations will satisfy the following field equations for the Levi-Civita connection Minkowski space background: 
\begin{align}
                \begin{split}
                        \lambda_0^2\left(\eta^{dk}\partial_d\partial_p\hat{\mathcal{P}}_{kq} + \eta^{dk}\partial_d\partial_q\hat{\mathcal{P}}_{kp} - 2\eta^{dk}\partial_d\partial_k\hat{\mathcal{P}}_{pq}\right)\\
                        - \left(d-1\right)^2\eta_{pq}\eta^{ab}\hat{\mathcal{P}}_{ab} + 2\left(2d-3\right)\left(d-1\right)\hat{\mathcal{P}}_{pq} = 0.
                \end{split}
        \end{align}
        
        \section{$\mathcal{P}_{ab}$ Decomposition}
        
        The decomposition of $\mathcal{P}_{ab}$ is inspired by the generic decomposition of symmetric rank 2 tensor discussed by York \cite{York}. The solution space for the $\hat{\mathcal{P}}_{ab}$ fluctuations can be neatly categorized by examining the irreducible representations of the Fourier transform $\tilde{\hat{\mathcal{P}}}_{ab}.$ $\tilde{\hat{\mathcal{P}}}_{ab}$ is a symmetric rank-2 tensor field in d=4 Minkowski space in Cartesian coordinates. This can be first decomposed into trace and traceless parts: 
\begin{align}
                \hat{\mathcal{P}}_{ab} = \left(\hat{\mathcal{P}}_{0}\right)_{ab} + \tfrac{1}{d}\hat{\mathcal{P}}\eta_{ab}, 
        \end{align}
corresponding to a $10 = 9 \oplus 1$ decomposition.  The traceless part can be further decomposed into transverse and longitudinal w.r.t the wavevector $k^a$:
\begin{align}
                &\left(\hat{\mathcal{P}}_{0}\right)_{ab}=\left(\hat{\mathcal{P}}_{TT}\right)_{ab} + \left(\hat{\mathcal{P}}_{LT}\right)_{ab}\\
                &\nabla^a\left(\hat{\mathcal{P}}_{TT}\right)_{ab} = 0  \,\Rightarrow \, k^a\left(\tilde{\hat{\mathcal{P}}}_{TT}\right)_{ab} = 0,
        \end{align}
further reducing the traceless component to $9 = 5 \oplus 4.$ Furthermore, the longitudinal part, $\left(\hat{\mathcal{P}}_{LT}\right)_{ab}$, can be expressed in terms of a 4-vector $L_b$,
\begin{align}
                &L_b = \nabla^a\left(\hat{\mathcal{P}}_{0}\right)_{ab} = 0 \, \Rightarrow \, k^a\left(\tilde{\hat{\mathcal{P}}}_{0}\right)_{ab} =0,\cr& \text{and} \, \nabla^a\left(\hat{\mathcal{P}}_{LT}\right)_{ab} =0 \, \Rightarrow k^a\left(\tilde{\hat{\mathcal{P}}}_{LT}\right)_{ab}=0,
        \end{align}
with $\left(\hat{\mathcal{P}}_{LT}\right)_{ab}$ traceless, namely, $\eta^{ab}\left(\hat{\mathcal{P}}_{LT}\right)_{ab} = 0$ and 
\begin{align}
                \left(\hat{\mathcal{P}}_{LT}\right)_{ab} = \tfrac{1}{2} k_{(a}L_{b)} - \tfrac{1}{d}\eta_{ab} k^cL_c.
        \end{align}
         
Now, consider how the 4-vector $L_a$ can also be decomposed w.r.t. $k^a$: 
\begin{align}
                L_a = \left(L_T\right)_a + \left(L_l\right)_a;~~& k^a\left(L_T\right)_a = 0;~~ S_L = k^a\left(L_l\right)a\\
                &\left(L_l\right)_a = S_L k^{-2}k_a
        \end{align}
where $\left(L_T\right)_a = L_a - S_L k^{-2}k_a$. Now we can relate $L_a$ to the degrees of freedom of $S_L$ and $\left(L_T\right)_a$: $L_a = \tfrac{d-1}{d}S_Lk_a + \tfrac{1}{2}k^2\left(L_T\right)_a$ giving a further reduction of  $4 = 3 \oplus 1.$ Returning to the decomposition of $\hat{\mathcal{P}}_{ab}$ into trace and traceless parts, we have the reducible decomposition of  $10 = 5 \oplus 4 \oplus 1$,
 \begin{align}
                \tilde{\hat{\mathcal{P}}}_{ab} = \left(\tilde{\hat{\mathcal{P}}}_{TT}\right)_{ab} + \left(\tilde{\hat{\mathcal{P}}}_{LT}\right)_{ab} + \tfrac{1}{d}\tilde{\hat{\mathcal{P}}}\eta_{ab},
        \end{align}
with,        
        \begin{align}
                &L_b = \left(L_{\text{Sol}}\right)_b + \left(L_{\text{Scalar}}\right)_b,\\
                &\nabla^b\left(L_{\text{Sol}}\right)_b = k^b\left(\tilde{L}_{\text{Sol}}\right)_b = 0, \,\text{and}\\
                &\nabla^b\left(L_{\text{Scalar}}\right)_b = S_L \,\Rightarrow~~ k^b\left(\tilde{L}_{\text{Scalar}}\right)_b = \tilde{S_L.}
        \end{align}
 The longitudinal-traceless part, $\left(\tilde{\hat{\mathcal{P}}}_{LT}\right)_{ab}$, of $\tilde{\hat{\mathcal{P}}}_{ab}$ can be further decomposed into the scalar and solenoidal parts,
\begin{align}
                \left(\tilde{\hat{\mathcal{P}}}_{LT}\right)_{ab} = \left(\tilde{\hat{\mathcal{P}}}_{\text{Sol}}\right)_{ab} + \left(\tilde{\hat{\mathcal{P}}}_{\text{Scalar}}\right)_{ab},
        \end{align}
 where
\begin{align}
                \left(\tilde{\hat{\mathcal{P}}}_{\text{Sol}}\right)_{ab} &= \tfrac{1}{2}\left[k_{(a}\left(L_T\right)_{b)}\right] \, \text{and}\\
                \left(\tilde{\hat{\mathcal{P}}}_{\text{Scalar}}\right)_{ab} &= S_L\left(k^{-2}k_ak_b - \tfrac{1}{d}\eta_{ab}\right).
        \end{align}
Putting it all together, the irreducible decomposition of $\tilde{\hat{\mathcal{P}}}_{ab}$ is then $10 = 5 \oplus 3 \oplus 1 \oplus 1:$
        
        \begin{align}
                \tilde{\hat{\mathcal{P}}}_{ab} = \left(\tilde{\hat{\mathcal{P}}}_{TT}\right)_{ab} + \left(\tilde{\hat{\mathcal{P}}}_{\text{Sol}}\right)_{ab} + \left(\tilde{\hat{\mathcal{P}}}_{\text{Scalar}}\right)_{ab} + \tfrac{1}{d}\tilde{\hat{\mathcal{P}}}\eta_{ab}.
        \end{align}
The transverse-traceless sector abides by the following identities:
\begin{align}
                \eta^{ab}\left(\hat{\mathcal{P}}_{TT}\right)_{ab} = \eta^{ab}\left(\tilde{\hat{\mathcal{P}}}_{TT}\right)_{ab} = 0\\
                \nabla^b\left(\hat{\mathcal{P}}_{TT}\right)_{ab} = k^b\left(\tilde{\hat{\mathcal{P}}}_{TT}\right)_{ab} = 0,
        \end{align}
whereas the longitudinal-traceless-solenoidal sector abides by
\begin{align}
                \eta^{ab}\left(\hat{\mathcal{P}}_{\text{Sol}}\right)_{ab} = \eta^{ab}\left(\tilde{\hat{\mathcal{P}}}_{\text{Sol}}\right)_{ab} = 0\\
                \nabla^a\nabla^b\left(\hat{\mathcal{P}}_{\text{Sol}}\right)_{ab} = k^ak^b\left(\tilde{\hat{\mathcal{P}}}_{\text{Sol}}\right)_{ab} = 0.
        \end{align}
 Using these identities, projection operators that project out the irreducible components  of   $\tilde{\hat{\mathcal{P}}}_{ab}$ can be constructed. To begin, we define the projection operator, $\hat{T_0}$, that extracts the traceless component. Thus we have the scalar $\mathcal{P}$ and the traceless tensor   $(\hat{\mathcal{P}}_{0})_{ab}$ defined through,
\begin{align}
                &\hat{\mathcal{P}} = \hat{\mathcal{P}}_{ab}\eta^{ab} \, \text{and}\\
                &\left(\hat{\mathcal{P}}_{0}\right)_{ab} = \left(\hat{T_0}\right)^{ij}{}_{ab}\left(\hat{\mathcal{P}}\right)_{ij}\, \text{respectively with}\\
                &\left(\hat{T_0}\right)^{ij}{}_{ab} = \delta^i{}_{a}\delta^j{}_{b} - \tfrac{1}{d}\eta_{ab}\eta^{ij}.
        \end{align}

Next we decompose the traceless sector. First the traceless-longitudinal-scalar projection operator, $\left(\tilde{\hat{LT}}_{\text{Scalar}}\right)^{ij}{}_{ab}$  is constructed:
\begin{align}
                &\left(\tilde{\hat{\mathcal{P}}}_{\text{Scalar}}\right)_{ab} = \left(\tilde{\hat{LT}}_{\text{Scalar}}\right)^{ij}{}_{ab}\tilde{\hat{\mathcal{P}}}_{ij} \, \text{with}\\
                &\left(\tilde{\hat{LT}}_{\text{Scalar}}\right)^{ij}{}_{ab} = \tfrac{d}{d-1}k^{-2}k^uk^v\left(\tilde{\hat{T_0}}\right)^{ij}{}_{uv}\left(k^{-2}k_ak_b - \tfrac{1}{d}\eta_{ab}\right)
        \end{align}
Then traceless-longitudinal-solenoidal sector is recovered from removing the traceless-longitudinal-scalar sector from the traceless-longitudinal sector, so that:
\begin{align}
                &\left(\tilde{\hat{\mathcal{P}}}_{\text{Sol}}\right)_{ab} = \left(\tilde{\hat{LT}}_{\text{Sol}}\right)^{ij}{}_{ab}\tilde{\hat{\mathcal{P}}}_{ij} \, \text{with},\\
                &\left(\tilde{\hat{LT}}_{\text{Sol}}\right)^{ij}{}_{ab} = \tfrac{1}{2}\left[k_{(a}\left(\tilde{\hat{L}}_T\right)^{ij}{}_{b)}\right] \,\,\text{and}\\
                &\left(\tilde{\hat{L}}_T\right)^{ij}{}_{a} = 2k^{-2}k^b\left[\left(\tilde{\hat{T_0}}\right)^{ij}{}_{ab} - \left(\tilde{\hat{LT}}_{\text{Scalar}}\right)^{ij}{}_{ab}\right]
        \end{align}
as the projectors.  With the traceless-longitudinal sector identified, the transverse-traceless sector is separated from the traceless sector by removing the traceless-longitudinal-solenoidal and traceless-longitudinal-scalar sectors.
This is achieved via the projection operators defined through,         
        \begin{align}
                &\left(\tilde{\hat{\mathcal{P}}}_{TT}\right)_{ab} = \left(\tilde{\hat{TT}}\right)^{ij}{}_{ab}\hat{\mathcal{P}}_{ij}\\
                &\left(\tilde{\hat{TT}}\right)^{ij}{}_{ab} = \left[\left(\tilde{\hat{T}}_0\right)^{ij}{}_{ab} - \left(\tilde{\hat{S}}_L\right)^{ij}{}_{ab} - \left(\tilde{\hat{LTS}}\right)^{ij}{}_{ab}\right]
        \end{align}
with these operators defined, $\tilde{\hat{\mathcal{P}}}_{ab}$ is decomposed into the irreducible components of $10 = 5 \oplus 3 \oplus 1 \oplus 1$. In what follows, we show that the transverse-longitudinal-scalar and the trace are related.
        
        \subsection{Scalar-Trace Condition from Field Equations}
The next condition is inspired by the transversality condition for the massive Proca equation; The Projective Cotton-York tensor is antisymmetric in two of its indices like the Faraday tensor. In Minkowski space with compatible connection, it's then killed when taking two divergence,
\begin{align}
                \bar{\nabla}_a\bar{\nabla}_b\bar{\nabla}_{c}K^{(ab)c} = \bar{\nabla}_a\left(\bar{R}^a{}_{mcb}K^{mbc}\right).
        \end{align}
Returning to the original diffeomorphism field equations, we take the connection to be compatible with Minkowski space and take two divergences resulting in
\begin{align}
                \lambda_0^2\nabla_i\nabla_j\nabla_{d}K^{(ij)d}&-\left(d-1\right)g^{ij}\nabla_i\nabla_j\mathcal{K}\cr
                &+2\left(2d-3\right)\nabla_i\nabla_j\mathcal{K}^{ij} = 0.
        \end{align}
For the problem in Minkowski space with compatible connection in Cartesian coordinates, the spacetime is Ricci Flat, and our curvature tensors for the fluctuation $\hat{\mathcal{P}}_{ab}$ become
\begin{align}
                -\left(d-1\right)\partial^i\partial_i\hat{\mathcal{P}}+2\left(2d-3\right)\partial^i\partial^j\hat{\mathcal{P}}_{ij} = 0.
        \end{align}
The fluctuation can the be expanded into its irreducible representations, and recall that the transverse-traceless sector is divergenceless and the solenoidal sector is double divergenceless,\begin{align}
                \begin{split}
                        &-\left(d-1\right)\partial^i\partial_i\hat{\mathcal{P}}\\
                        &+2\left(2d-3\right)\partial^i\partial^j\left(\left(\hat{\mathcal{P}}_{LT_{\text{Scalar}}}\right)_{ij} + \tfrac{1}{d}\hat{\mathcal{P}}\eta_{ij}\right) = 0.
                \end{split}
        \end{align}
The transverse-longitudinal-solenoidal sector vanishes under two divergences and is omitted. Taking the Fourier transform and collecting the like terms, the field equations give a relationship between the trace and transverse-longitudinal-scalar sectors,
\begin{align}
                \partial^a\partial^b\left(\hat{\mathcal{P}}_{LT_{\text{Scalar}}}\right)_{ab} &= \left(\tfrac{\left(d-2\right)\left(d-3\right)}{2d\left(2d-3\right)}\right)\partial^i\partial_i\hat{\mathcal{P}}\cr
                k^ak^b\left(\tilde{\hat{\mathcal{P}}}_{LT_{\text{Scalar}}}\right)_{ab} &= \left(\tfrac{\left(d-2\right)\left(d-3\right)}{2d\left(2d-3\right)}\right)k^2\tilde{\hat{\mathcal{P}}}.
        \end{align}
The transverse-longitudinal-scalar sector has a single degree of freedom which gives the relationship
\begin{align}
                k^ak^b\left(\tilde{\hat{\mathcal{P}}}_{LT_{\text{Scalar}}}\right)_{ab} = \left(\tfrac{d-1}{d}\right)k^2\tilde{S}_L.
        \end{align}
The Diffeomorphism field equations then relates the two scalar degrees of freedom via,
\begin{align}
                \tilde{S}_L = \left(\tfrac{\left(d-2\right)\left(d-3\right)}{2\left(d-1\right)\left(2d-3\right)}\right)\tilde{\hat{\mathcal{P}}}
        \end{align}
In the case of 4-dimensional spacetime $\tilde{S}_L = \tfrac{1}{15}\tilde{\hat{\mathcal{P}}}$.
        
        \subsection{Decomposed Field Equations}

Using the scalar-trace condition, the diffeomorphism field fluctuations, $\hat{\mathcal{P}}_{ab}$, has the full decomposition:
\begin{align}
                \begin{split}
                        \hat{\mathcal{P}}_{ab} = &\left(\hat{\mathcal{P}}_{TT}\right)_{ab} + \left(\hat{\mathcal{P}}_{LT}\right)_{ab}\\
                        & + \left(\tfrac{1}{d}\eta_{ab} + \left(\tfrac{\left(d-2\right)\left(d-3\right)}{2\left(d-1\right)\left(2d-3\right)}\right)\left(k^{-2}k_ak_b - \tfrac{1}{d}\eta_{ab}\right)\right)\hat{\mathcal{P}},
                \end{split}
        \end{align}
where the 10 degrees of freedom $\hat{\mathcal{P}}_{ab}$ might have possessed from being a symmetric rank 2 tensor is reduced to 9 from the scalar-trace condition giving $9 = 5 \oplus 3 \oplus 1.$ The transverse-traceless projection operator within the specifications of the scalar-trace condition becomes
 \begin{align}
                \begin{split}
                        \left(\hat{TT}\right)^{ij}{}_{ab} = &\left(\hat{T}_0\right)^{ij}{}_{ab} - \left(\hat{LT}\right)^{ij}{}_{ab}\\
                        & - \left(\tfrac{\left(d-2\right)\left(d-3\right)}{2\left(d-1\right)\left(2d-3\right)}\right)\left(k^{-2}k_ak_b - \tfrac{1}{d}\eta_{ab}\right)\eta^{ij}.\\
                \end{split}
        \end{align}
Whereas the $\left(\hat{L_T}\right)$ projection operator becomes
        
        \begin{align}
                \begin{split}
                        \left(\hat{L}_T\right)^{ij}{}_{a}& = \\
                        2k^{-2}k^b&\left[\left(\hat{T}_0\right)^{ij}{}_{ab} - \left(\tfrac{\left(d-2\right)\left(d-3\right)}{2\left(d-1\right)\left(2d-3\right)}\right)\left(k^{-2}k_ak_b - \tfrac{1}{d}\eta_{ab}\right)\eta^{ij}\right]
                \end{split}
        \end{align}
With these operators in the context of the scalar-trace condition, the decomposition of $\hat{\mathcal{P}}_{ab}$ is now,
\begin{align}
                \begin{split}
                        \tilde{\hat{\mathcal{P}}}_{ab} = &\left(\tilde{\hat{\mathcal{P}}}_{TT}\right)_{ab} + \left(\tilde{\hat{\mathcal{P}}}_{\text{Sol}}\right)_{ab}\\
                        & + \left(\tfrac{1}{d}\eta_{ab} + \left(\tfrac{(d-2)(d-3)}{(d-1)(4d-6)}\right)\left(k^{-2}k_ak_b - \tfrac{1}{d}\eta_{ab}\right)\right)\tilde{\hat{\mathcal{P}}}.
                \end{split}
        \end{align}
The diffeomorphism fluctuation field equations reduce to the following for the respective sectors:
\begin{align}
                \left(\square - \tfrac{\left(d-1\right)\left(2d-3\right)}{\lambda_0^2}\right)\left(\hat{\mathcal{P}}_{TT}\right)_{ab} &= 0,\\
                \left(\square - \tfrac{2\left(d-1\right)\left(2d-3\right)}{\lambda_0^2}\right)\left(\hat{\mathcal{P}}_{LT}\right)_{ab} &= 0,\\
                \left(\square + \tfrac{d(d-2)(d-3)}{2\lambda_0^2\left(1-\tfrac{(d-2)(d-3)}{2(d-1)(2d-3)}\right)}\right)\hat{\mathcal{P}} &= 0.
        \end{align}
When interpreted as free fields, the scalar becomes lightlike in $d<4$ dimensions. In dimensions $d<3$ there are no transverse traceless degrees of freedom. The mass squared terms appearing in the scalar field equations match the poles appearing in the generic green function for the diffeomorphism field. Choosing $d=4$ for spacetime dimensions the field equations become,
\begin{align}
                \left(\square - \tfrac{15}{\lambda_0^2}\right)\left(\hat{\mathcal{P}}_{TT}\right)_{ab} &= 0,\\
                \left(\square - \tfrac{30}{\lambda_0^2}\right)\left(\hat{\mathcal{P}}_{LT}\right)_{ab} &= 0,\\
                \left(\square + \tfrac{30}{7\lambda_0^2}\right)\hat{\mathcal{P}} &= 0.
        \end{align}
        
        This decomposition shows that the trace is a massive field where as the transverse-traceless and transverse-longitudinal-solenoidal sectors are tachyonic fields with different tachyonic masses. As discussed in Ref.\cite{GS}, the scalar component is the only component that couples directly to matter fields (fermions).  All of the masses are inversely proportional to the projective length scale $\lambda_0.$ The mass terms present in the respective field equations endow each of the species with their own dispersion relationship $\omega = \sqrt{k^2 + m^2},$ where $\omega$ is the time component of a generalized 4-wavevector, $k^a,$ and $k^2$ is the length squared of the 3-wavevector. The generic 4-wavevector is then, $k_{a} = \left(k_x,k_y,k_z,\omega\left(\vec{k}\right)\right).$ The scalar wave for fixed wavenumber can be written as
\begin{align}
                \left(\hat{\mathcal{P}}_{\text{Scalar}}\right)_{ab}(x^u) = \mathcal{P}\left(\vec{k}\right)\left[\eta_{ab} + \tfrac{\lambda_0^2}{15}k_ak_b\right]e^{ik_ux^u}.
        \end{align}
        
        The $\mathcal{P}\left(\vec{k}\right)$ being the scalar amplitude degree of freedom for fixed wavenumber. A generalized wavepacket can be assembled by integrating over $\vec{k}.$ The solenoidal wave for fixed wavenumber is
        
        \begin{align}
                &\left(\hat{\mathcal{P}}_{\text{Sol}}\right)_{ab}(x^u) = \tfrac{1}{2}\left[k_{(a}\left(L_T\right)_{b)}\right]e^{ik_ux^u}\\
                &k^a\left(L_T\right)_{a} = 0.
        \end{align}
The transversality condition on $\left(L_T\right)_{a}$ reminds us that it only has 3 degrees of freedom. A generalized wavepacket for the solenoidal wave can be formed from integrating $\left(\hat{\mathcal{P}}_{LT_\text{Sol}}\right)_{ab}$ over $\vec{k}.$ For the transverse-traceless waves, $\left(\tilde{\hat{\mathcal{P}}}_{TT}\right)_{ab} = \left(\tilde{\hat{TT}}\right)^{ij}{}_{ab}\hat{\mathcal{P}}_{ij}$ can be used where $\hat{\mathcal{P}}_{ij}$ is expressed as a generic symmetric rank 2 tensor with specific wavenumber $\vec{k}$ and $\left(\tilde{\hat{TT}}\right)^{ij}{}_{ab}$ with specific wavenumber $\vec{k}.$ $\left(\tilde{\hat{\mathcal{P}}}_{TT}\right)_{ab}$ components are then determined by 5 parameters. As an example we can take wavevector $\vec{k} = \left(0,0,k\right),$ then $\left(\tilde{\hat{\mathcal{P}}}_{TT}\right)_{ab}$ is
        
        \begin{align}
                \begin{split}
                        &\left(\hat{\mathcal{P}}_{TT}\right)_{ab} = e^{ik_ux^u}\times\\
                        &\begin{bmatrix}
                                \mathcal{P}_{xx} & \mathcal{P}_{xy} & \mathcal{P}_{xz} & \tfrac{k\mathcal{P}_{xz}}{\sqrt{k^2 -\tfrac{15}{\lambda_0^2}}}\\
                                \mathcal{P}_{xy} & -\mathcal{P}_{xx} + \tfrac{15\mathcal{P}_{zt}}{k\lambda_0^2\sqrt{k^2 -\tfrac{15}{\lambda_0^2}}} & \mathcal{P}_{yz} & \tfrac{k\mathcal{P}_{yz}}{\sqrt{k^2 -\tfrac{15}{\lambda_0^2}}}\\
                                \mathcal{P}_{xz} & \mathcal{P}_{yz} & \tfrac{\mathcal{P}_{zt}\sqrt{k^2 -\tfrac{15}{\lambda_0^2}}}{k} & \mathcal{P}_{zt}\\
                                \tfrac{k\mathcal{P}_{xz}}{\sqrt{k^2 -\tfrac{15}{\lambda_0^2}}} & \tfrac{k\mathcal{P}_{yz}}{\sqrt{k^2 -\tfrac{15}{\lambda_0^2}}} & \mathcal{P}_{zt} & \tfrac{k\mathcal{P}_{zt}}{\sqrt{k^2 -\tfrac{15}{\lambda_0^2}}}
                        \end{bmatrix}.
                \end{split}
        \end{align}
        
        \noindent The transverse-traceless wave is then specified in this instance by the five parameters: $\mathcal{P}_{xx},$ $\mathcal{P}_{xy},$ $\mathcal{P}_{xz},$ $\mathcal{P}_{yz},$ and $\mathcal{P}_{zt}.$
        
        With the full decomposition for $\hat{\mathcal{P}}$ in hand, we can revisit the TW action. Expressing the TW action in a Minkowski space and Levi-Civita background, and expanding the diffeomorphism field about the ``expectation value'' $\mu = \tfrac{-1}{8J_0\kappa_0c}$ and massive transverse-traceless, solenoidal, and trace fields the TW action becomes
\begin{align}
                &\mathcal{L}_{\text{Mink}} =&\cr
                J_0c\Big[-2\Big(&\lambda_0^2\left(\mathcal{P}_{TT}\right)_{ab}\Box\left(\mathcal{P}_{TT}\right)^{ab}\cr
                & - \left(d-1\right)\left(2d-3\right)\left(\mathcal{P}_{TT}\right)_{ab}\left(\mathcal{P}_{TT}\right)^{ab}\Big)\cr
                - \Big(&\lambda_0^2\left(\mathcal{P}_{\text{Sol}}\right)_{ab}\Box\left(\mathcal{P}_{\text{Sol}}\right)^{ab}\cr
                &-2\left(d-1\right)\left(2d-3\right)\left(\mathcal{P}_{\text{Sol}}\right)_{ab}\left(\mathcal{P}_{\text{Sol}}\right)^{ab}\Big)\cr
                -\Big(&\lambda_0^2\left(\tfrac{(3d-5)^2}{2(2d-3)^2(d-1)}\right)\hat{\mathcal{P}}\Box\hat{\mathcal{P}}\cr
                & + \left(\tfrac{3d^3 - 20d^2 + 43d-30}{2(2d-3)}\right)\hat{\mathcal{P}}^2\Big)\Big] + \dots.
        \end{align}
\noindent The TW action then describes three decoupled  free fields  for the diffeomorphism field in a Minkowski space with  Levi-Civita connection.
        
        \subsection{Tachyon Considerations}
        
        The transverse-traceless and solenoidal-traceless fields have imaginary masses when interpreted as as free fields making them tachyonic. The presence of tachyonic mass terms indicate an instability where the transverse-traceless and solenoidal-traceless fields either exponentially grow or decay \cite{Sus}. However, we can consider two aspects of measuring their presence: the decomposition of the $\mathcal{P} $-field into it's irreducible representations w.r.t. $k^a$ and $\eta_{ab}$ have diagonalized the kinetic terms for the $\mathcal{P}$-field. After introducing perturbations, $h_{ab},$ in the metric,  $\mathcal{P}_{ab}$ and $h_{ab}$ will have contact terms.  After diagonalization, the tachyonic modes could be interpreted as resonances in graviton-graviton scattering. Secondly, in an early universe epoch, there could have been an initial non-zero tachyonic field contribution that assisted in driving inflation as well as  sourcing traditional linearized metric gravity waves. These modes could have   evanescently decayed, ending inflation.   
Earlier work has shown that the scalar component in an FRWL geometry serves as a candidate for the inflaton \cite{Abdullah:2022dzz}.         
%
%
%
%
        
        \section{Geodesic Deviation}
        
        We can now consider the geodesic deviation from our fluctuation $\hat{\mathcal{P}}_{ab}.$ This geodesic deviation is about fiducial geodesics, solutions to the projective geodesic equation with projective connections given by the Minkowski space background solution.
        
        \subsection{Geodesics}
        
        The Thomas Cone geodesics are solutions to the projective geodesic equation,
\begin{align}
                \frac{d^2x^\alpha}{du^2} + \tilde{\Gamma}^\alpha{}_{\mu\nu}\tfrac{dx^\mu}{du}\tfrac{dx^\nu}{du} &= 0,
        \end{align}
with connection components $\tilde{\Gamma}^\alpha{}_{\mu\nu}$ (see Eq.\ref{eq: Gammatilde}) consistent with the Minkowski space solution. Here $u$ is an arbitrary parameter.  This can be expressed in terms of $\Pi$ and $\cD$, splitting into the usual  manifold geodesic equations  and another expression for the $\l$coordinate.  That equation is a  Sturm-Liouville equation with a potential given by $D_{u u} \equiv \mathcal{D}_{bc}\tfrac{dx^b}{du}\tfrac{dx^c}{du}$, which is the pullback of the diffeomorphism field onto the geodesics line parameterized by $u$.  The two set of equations are:\begin{align}
                \tfrac{d^2x^a}{du^2} + \Pi^a{}_{bc}\tfrac{dx^b}{du}\tfrac{dx^c}{du} &= -2\tfrac{1}{\lambda}\left(\tfrac{d\lambda}{du}\right)\tfrac{dx^a}{du}\\
                \tfrac{d^2\lambda}{du^2} + \lambda\mathcal{D}_{bc}\tfrac{dx^b}{du}\tfrac{dx^c}{du} &= 0.
        \end{align}
        
For the Minkowski Space solutions, with $\Pi^a{}_{bc}=0$ and $\cD_{ab} = -\tfrac{1}{8J_0\kappa_0c}\eta_{ab}$, the projective geodesic equations become
\begin{align}
                &\tfrac{d^2x^a}{du^2} = -2\tfrac{1}{\lambda}\left(\tfrac{d\lambda}{du}\right)\tfrac{dx^a}{du},\\
                &\tfrac{d^2\lambda}{du^2} - \tfrac{\lambda}{8J_0\kappa_0c}\eta_{bc}\tfrac{dx^b}{du}\tfrac{dx^c}{du} = 0.
        \end{align}
One sees, from the first equation,  that the logarithmic derivative of $\l  $ with respect to the parameter $u$ acts as a ``geometrical friction'' force on particles following along this geodesics, viz.  $\tfrac{1}{\lambda}\left(\tfrac{d\lambda}{du}\right)$ .  The Sturm-Liouville equation described the nature of this force that has become manifest due to $D_{u u} $ not being zero.    Had $D_{u u} =0 $, then $\l = a u +b$, and the ``geometric friction" can be eliminated by rescaling the vector fields.    

The manifold geodesic equations are separable and can be recognized as a relationship between logarithmic derivatives,        
        \begin{align}
                \tfrac{d}{du}\log\left(\tfrac{dx^a}{du}\right) = \tfrac{d}{du}\log\left(\lambda^{-2}\right)
        \end{align}
 for each direction separately. Directly integrating gives $\tfrac{dx^a}{du}$ with integration constant $c^a,$
\begin{align}
                \tfrac{dx^a}{du} = \tfrac{c^a}{\lambda^2}.
        \end{align}
 Now, we can choose $c^a = \left(1,0,0,0\right)$ for a fiducial time-like geodesic provided that $\eta_{bc}\tfrac{dx^b}{du}\tfrac{dx^c}{du}> 0$. The Sturm-Liouville equation is then
 \begin{align}
                \tfrac{d^2\lambda}{du^2} = \frac{c}{8J_0\kappa_0}\left(\tfrac{1}{\lambda^3}\right).
        \end{align}
For convenience, define $A = \sqrt{\tfrac{8J_0\kappa_0}{c}}.$ The general solution for $\lambda$ depends on two coefficients $c_1$ and $c_2, $ namely
        \begin{align}
                \lambda = \pm A^{-1}\sqrt{\tfrac{1+A^2c_1^2\left(u+c_2\right)^2}{c_1}}.
        \end{align}
 The $\lambda$ component of the Thomas cone being nonnegative restricts us to the positive solutions. Returning to the manifold equations, only the time coordinate remains relevant with our chosen initial conditions. The time coordinate in our initial parameterization is labeled $x^0=t_1.$ Therefore we can directly integrate the time geodesic equation as,
\begin{align}
                \tfrac{dx^0}{du} &=  \tfrac{Ac_1}{1+A^2c_1^2\left(u+c_2\right)^2}\\ \, \text{which yields} \,\,&\\
                t_1 &= A\arctan\left[Ac_1\left(u+c_2\right)\right] + c_3.         \end{align}
        
        Now we perform a transformation of the coordinates on the Thomas cone (see \ref{TCTransformations})   from $x^\alpha = \{0,0,0,t_1,\lambda_1\} \rightarrow x'^\alpha = \{0,0,0,t_2,\lambda_2\}$ so that $t_2=c_1\left(u+c_2\right)$, is  an affine transformation and $u$ can be interpreted as proper time $\t$. Then the transformation,   \begin{equation}
               t_2 = A^{-1}\tan\left[A^{-1}\left(t_1-c_3\right)\right] = c_1\left(\tau+c_2\right),
\end{equation} with
\begin{equation}\lambda_1 = A^{-1}\sqrt{\tfrac{1+A^2c_1^2\left(\tau+c_2\right)^2}{c_1}},\end{equation}we find that the corresponding  $\l$ is
\begin{equation} \lambda_2 = \lambda_1 |\boldsymbol{J}|^{-\tfrac{1}{5}} = A^{}\left(A^{-2} + c_1^2\left(\tau+c_2\right)^2\right)^{\tfrac{4}{5}}.\end{equation}
Then $\l$ would provide a friction force since     
\begin{equation}
\tfrac{d}{d\t}\log\left(\lambda\right) =\frac{8}{5} \frac{c_1^2 (c_2+ \tau)}{\frac{1}{A^2} + c_1^2(c_2+\tau)^2}.
\end{equation}   This demonstrates one of the salient features of a non-trivial diffeomorphism field on the cosmology and the  geometry and how it can modify geodesics thus  possibly mirroring dark matter.   
        
\subsection{Geodesic Deviation}
        
        At this stage we now consider the geodesic deviations about these fiducial geodesics. The geodesic deviation on the Thomas cone follows the usual derivation of geodesic deviation with connection components promoted to $\tilde{\Gamma}$, the connection on the Thomas cone. The usual manifold Riemann curvature tensor is promoted to the projective curvature tensor $\mathcal{K}.$ The geodesic deviation equation for deviation $\xi^A$ (the Jacobi field) about trajectory $\tfrac{dx^A}{d\tau}$ is then
\begin{align}
                \tfrac{D^2\xi^A}{d\tau^2}& + \mathcal{K}^{A}{}_{BCD}\tfrac{dx^{B}}{d\tau}\tfrac{dx^{D}}{d\tau}\xi^C = 0.
        \end{align}
        
        Evaluating the intrinsic derivative and exposing the connection components in $\mathcal{K}$ the geodesic deviation equation becomes
\begin{align}
                \tfrac{d^2\xi^\alpha}{d\tau^2}& + \left(\partial_\mu\tilde{\Gamma}^\alpha{}_{\rho\sigma}\right)\tfrac{dx^\rho}{d\tau}\tfrac{dx^\sigma}{d\tau}\xi^\mu + 2\tilde{\Gamma}^\alpha{}_{\rho\sigma}\tfrac{dx^\rho}{d\tau}\tfrac{d\xi^\sigma}{d\tau} = 0.
        \end{align}
When the connection components are specified by the  diffeomorphism field in the  the Minkowski space background, the deviation caused by any of the three distinct sectors of the diffeomorphism field becomes vacuous.  This follows since if we choose the initial conditions $\xi^{\lambda} = \tfrac{d\xi^{\lambda}}{d\tau} = 0$, the $\lambda$ equation of the geodesic deviation equations is trivially satisfied, and the remaining equations are simply
 \begin{align}
                \tfrac{2c_1^2\tau}{1+c_1^2\tau^2}\tfrac{d\xi^i}{d\tau} + \tfrac{d^2\xi^i}{d\tau^2} &= 0.
        \end{align}
        The geodesic deviation can then be specified by initial conditions $\xi^i(\tau=0) = b^i$ and $\tfrac{dx^i}{d\tau}(\tau=0) = b^i.$ After reparameterization the geodesic deviation is trivial.
 \begin{align}
                \xi^i &= \left(c_1\right)^{-1}a^i\arctan\left[c_1\tau\right]+b^i\\
                \xi'^i &= a^i\tau+b^i.
        \end{align}
The result here is that whatever deviation is measured will be based entirely of the initial conditions of the  geodesics. For example, the endpoints of the arms of an interferometer would initially just have a space-like separation with no velocity difference. The geodesic deviation would just be a constant separation between the arms without any influence from the diffeomorphism field. To $0^{th}$ order in metric fluctuations, the impact of the three diffeomorphism field species cannot be observed.
        
        \subsection{Approximate Metric Fluctuations}
        
        Now, we allow for limited back-reaction of the three diffeomorphism species by constructing their energy-momentum tensor as sources for linearized gravity. We consider first order corrections to Minkowski space, $h_{ab},$ such that
\begin{align}
                g_{ab} &\approx \eta_{ab} + h_{ab}\\
                g^{ab} &\approx \eta^{ab} - h^{ab}\\
                g_{ac}g^{cb} &= \delta_a{}^b.
        \end{align}
        
        For $T_{ab},$ the approximate energy momentum tensor constructed with $\eta$ such that $\left(T_{\text{Total}}\right)_{ab} = T_{ab} + \tau_{ab},$ with $\tau_{ab}$ is a pseudotensor that contains higher order contributions, the trace reversed Einstein field equations are sourced by
 \begin{align}
                S_{ab} = T_{ab} - \tfrac{1}{2}T\eta_{ab}.
        \end{align}
The first order fluctuation $h_{ab}$ can be gauge fixed by being put in harmonic gauge. The linearized metric field equations are then
\begin{align}
                \Box h_{ab} = 16\pi GS_{ab} .
        \end{align}
This is solved easily because of the convenient form of the source tensor $S_{ab}.$ The idealized fixed k-wavevector species provide factors of either $e^{ik_\mu x^\mu}$ or $e^{2ik_\mu x^\mu}$ to the energy-momentum tensor. $S_{ab}$ can then be decomposed into two polarization tensors, $\epsilon_1$ and $\epsilon_2$, for the respective pieces. The polarization tensor for fixed wavenumber have no spacetime coordinate dependency, so the Fourier transform, $\tilde{S}_{ab}$, is then
\begin{align}
                S_{ab} &=  \left(\epsilon_1\right)_{ab}e^{ik_\mu x^\mu} + \left(\epsilon_2\right)_{ab}e^{2ik_\mu x^\mu}\\
                \tilde{S}_{ab} &= \left(\epsilon_1\right)_{ab}\delta^{(4)}(k-q) + \left(\epsilon_2\right)_{ab}\delta^{(4)}(k-2q).
        \end{align}
The Fourier transform of the gauge fixed linearized metric equation is
\begin{align}
                -k^2 \tilde{h}_{ab} = 16\pi G\left[\left(\epsilon_1\right)_{ab}\delta^{(4)}(k-q) + \left(\epsilon_2\right)_{ab}\delta^{(4)}(k-2q)\right].
        \end{align}
By solving for $\tilde{h}_{ab}$ and taking the inverse Fourier transform we obtain
\begin{align}
                h_{ab} &=  \left(\epsilon_1\right)_{ab}e^{ik_\mu x^\mu} + \left(\epsilon_2\right)_{ab}e^{2ik_\mu x^\mu}.
        \end{align}
The polarizations $\epsilon_1$ and $\epsilon_2$ are rescaled by the mass-squared of the respective species, but can be absorbed into their magnitudes.
        
        \subsection{Numerical Geodesic Deviation}
        
        Equipped with the $h_{ab}$ field mentioned previously, its influence on geodesic deviation is examined without considering higher-order corrections in $h_{ab}$ or its backreaction in the diffeomorphism field equation for a higher order $S_{ab}.$ The geodesic deviation equations are sourced with the Thomas-cone connection featuring a nonvanishing $\Pi^a{}_{bc}$ constructed from $h_{ab}$ and diffeomorphism field that contains both the residual Minkowski space diffeomorphism contributions and the fluctuations of the respective species considered. The geodesic deviation equations are then solved numerically for the deviation $\xi$ about the time-like fiducial geodesic. For simplicity, we work with the natural timescale of $\sqrt{\tfrac{J_0\kappa_0}{c}}\rightarrow 1$ and diffeomorphism fluctuations traveling only in the z-direction so that $k_a = \{0,0,k,\sqrt{k^2+m^2}\}$, the $m^2$ being appropriate for the species considered. We choose as initial data for geodesic deviation, $\xi^\alpha(\tau=0) = \{\xi^x,\xi^y,0,0,0\}$ and $\dot{\xi}^\alpha(\tau=0) = 0.$ The choice to restrict initial data to deviation in the x-y plane is inspired by interferometers like LIGO where we take the arms of the interferometer to be in the x-y plane. Initial values $\xi^x$ and $\xi^y$ are chosen arbitrarily but with distinct values to differentiate on a graph. The magnitude of the diffeomorphism radiative degrees of freedom are exaggerated to demonstrate the salient features of the geodesic deviation. The choice of vanishing deviation components, $\xi^t$ and $\xi^\lambda$, are chosen for the separation of spatially separated geodesics to be simultaneous and on the same projective scaling $\lambda$. Present in all of the geodesic deviations is a nonzero $\xi^t$ component, which measures the difference in simultaneity for the remaining deviation components. $\xi^t\left(\tau\right)$ can then be interpreted as a calibration curve where the deviation needs to be considered at $\xi^\alpha\left(\tau'\right)$ where $\tau'= \tau-\xi^t(\tau).$
        
        \subsubsection{Scalar Deviation}
        
        The geodesic deviation for the scalar mode shows that spatial deviations perpendicular to the traveling wave is unaffected where as deviation in the z-direction gains a velocity. The time component of deviation indicating a difference in simultaneity of measuring the spatial deviation from the fiducial geodesic. The $\xi^\lambda$ component indicates the spatially separated geodesics with the same initial projective scaling $\lambda$ exponentially diverge. The deviation considered here is for a continuous scalar mode, but for a finite duration pulse, the geodesic deviation inherits the position and velocity it had at the end of the pulse duration. This can be considered a gravitational memory effect from diffeomorphism radiation. \cite{memory1}\cite{memory2} 
        
        \begin{figure}[h!]
                \caption{Example of Geodesic deviation of a time-like trajectory fiducial geodesic due to a scalar diffeomorphism fluctuation in the z-direction.}
                \centering
                \includegraphics[width=0.5\textwidth]{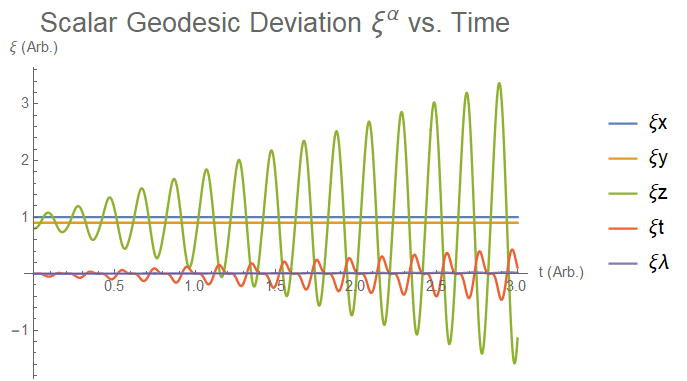}
        \end{figure}
        
        \subsubsection{Solenoidal Deviation}
        
        Similarly for solenoidal radiation, there are contributions to deviation in the $\xi^z,~ \xi^t,~\xi^\lambda$ components with an analogous gravitational memory effect. The solenoidal field, $\left(\mathcal{P}_{\text{Sol}}\right)_{ab} = \tfrac{1}{2}\left[k_{(a}\left(L_T\right)_{b)}\right]$, features three degrees of freedom from $\left(L_T\right)_a = \{P_1,P_2,P_3,\tfrac{kP_3}{\sqrt{k^2 - \tfrac{30}{\lambda_0^2}}}\}.$ Solenoidal radiation with nonzero $P_1$ and $P_2$ components produce changes in geodesic deviation in the $\xi^x$ and $\xi^y$ respectively.
        
        \begin{figure}[h!]
                \caption{Example of Geodesic deviation of a time-like trajectory fiducial geodesic due to a solenoidal diffeomorphism fluctuation in the z-direction.}
                \centering
                \includegraphics[width=0.5\textwidth]{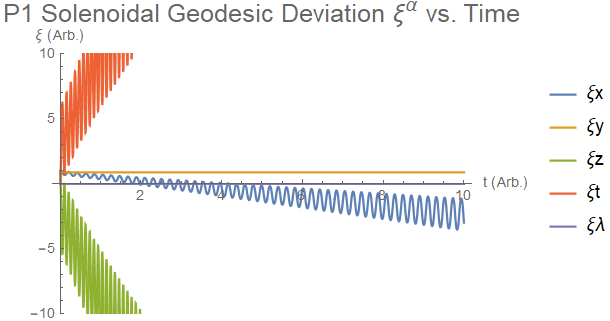}
        \end{figure}
        
        \begin{figure}[h!]
                \caption{Example of Geodesic deviation of a time-like trajectory fiducial geodesic due to a solenoidal diffeomorphism fluctuation in the z-direction.}
                \centering
                \includegraphics[width=0.5\textwidth]{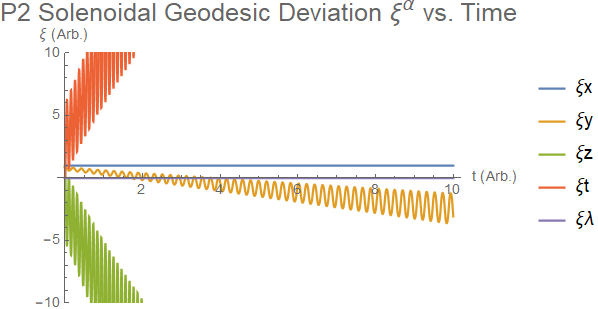}
        \end{figure}

        \begin{figure}[h!]
                \caption{Example of Geodesic deviation of a timelike trajectory fiducial geodesic due to a solenoidal diffeomorphism fluctuation in the z-direction.}
                \centering
                \includegraphics[width=0.5\textwidth]{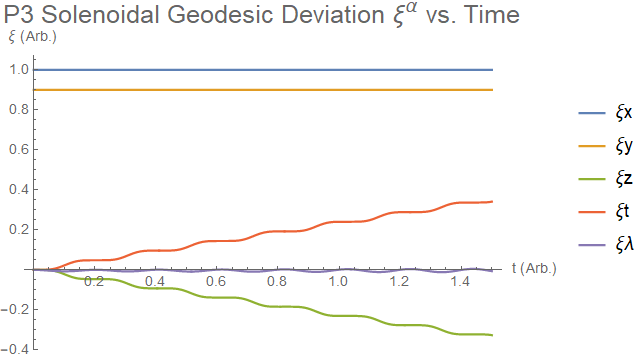}
        \end{figure}
        
        \subsubsection{Transverse Deviation}
        
        The transverse radiation continues with contributions to deviation in the $\xi^z,~ \xi^t,~\xi^\lambda$ components with an analogous gravitational memory effect. The $\left(\mathcal{P}_{\text{TT}}\right)_{xz}$ and $\left(\mathcal{P}_{\text{TT}}\right)_{yz}$ components contribute geodesic deviation in the $\xi^x$ and $\xi^y$ respectively.
        
        \begin{figure}[h!]
                \caption{Example of Geodesic deviation of a time-like trajectory fiducial geodesic due to a transverse diffeomorphism fluctuation in the z-direction.}
                \centering
                \includegraphics[width=0.5\textwidth]{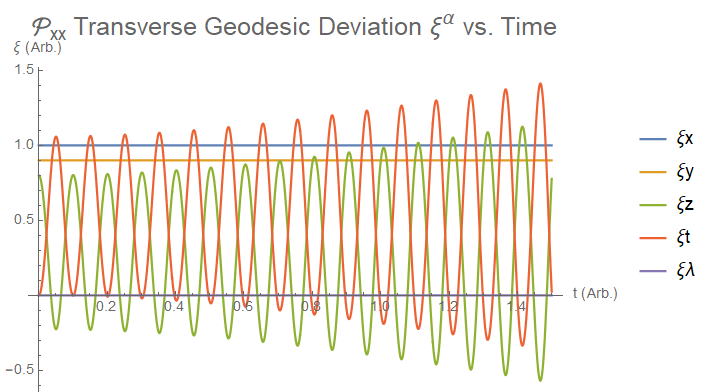}
        \end{figure}
        
        \begin{figure}[h!]
                \caption{Example of Geodesic deviation of a time-like trajectory fiducial geodesic due to a transverse diffeomorphism fluctuation in the z-direction.}
                \centering
                \includegraphics[width=0.5\textwidth]{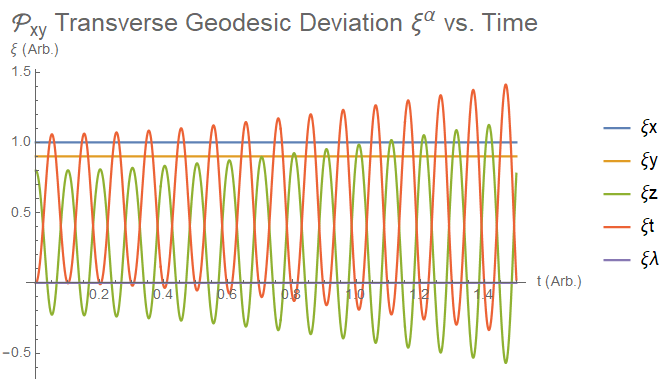}
        \end{figure}

        \begin{figure}[h!]
                \caption{Example of Geodesic deviation of a time-like trajectory fiducial geodesic due to a transverse diffeomorphism fluctuation in the z-direction.}
                \centering
                \includegraphics[width=0.5\textwidth]{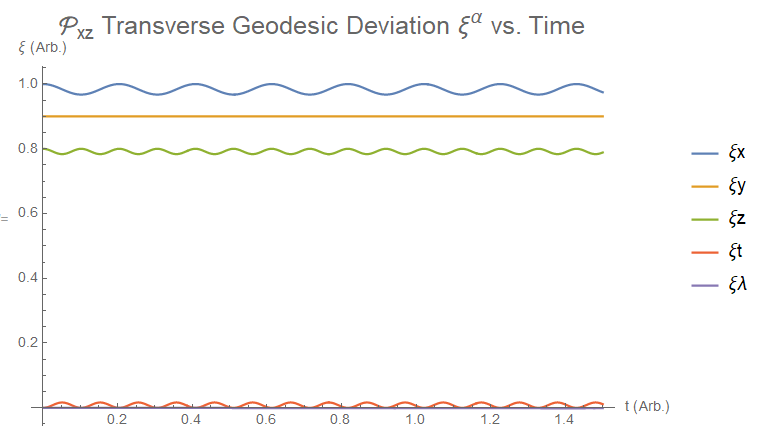}
        \end{figure}

        \begin{figure}[h!]
                \caption{Example of Geodesic deviation of a time-like trajectory fiducial geodesic due to a transverse diffeomorphism fluctuation in the z-direction.}
                \centering
                \includegraphics[width=0.5\textwidth]{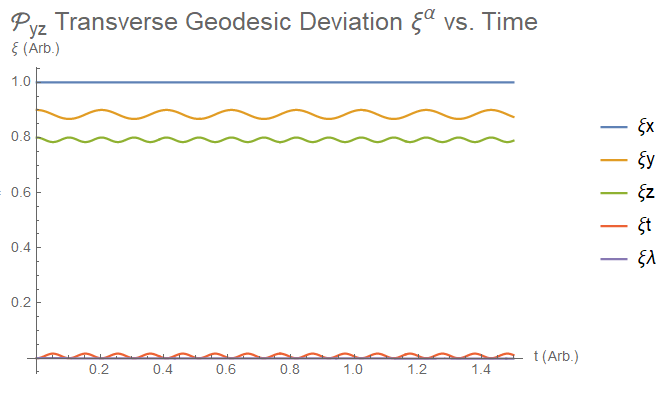}
        \end{figure}

        \begin{figure}[h!]
                \caption{Example of Geodesic deviation of a time-like trajectory fiducial geodesic due to a transverse diffeomorphism fluctuation in the z-direction.}
                \centering
                \includegraphics[width=0.5\textwidth]{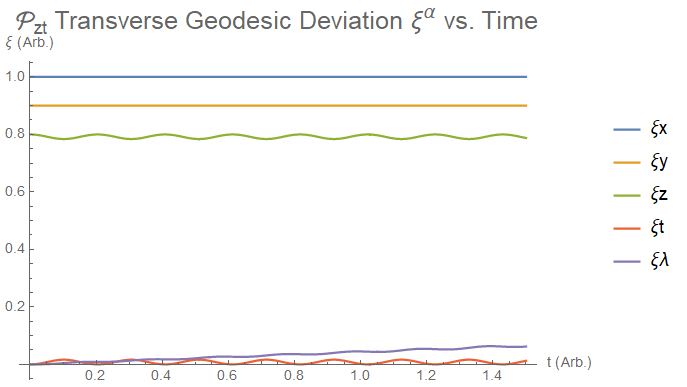}
        \end{figure}
        
        \subsection{Antenna Patterns}
        
        With the $h_{ab}$ from each of the respective species, the antenna sensitivity can be computed. The strategy of Forward is used to compute the antenna sensitivity for a gravitational Michelson Interferometer \cite{Forward}. For each polarization of tensor, vector, or scalar mode their is a corresponding $h_{ab}$ at first order. Consider the rotation matrix (in t,x,y,z ordering)
        \begin{align}
        	R^a{}_{b} = \begin{bmatrix}
        		1 & 0 & 0 & 0\\
        		0 & \cos\phi & \sin\phi & 0\\
        		0 & -\cos\theta\sin\phi & \cos\theta\cos\phi & \sin\theta\\
        		0 & \sin\theta\sin\phi & -\sin\theta\cos\phi & \cos\theta\\
        	\end{bmatrix}.
        \end{align}
        This matrix rotates a vector on the z axis to a new direction corresponding to a point on the unit sphere given in spherical coordinates with colatitude $\theta$ and azimuth $\phi$. The arbitrary inbound plane wave, $h'_{ab}\left(\theta,\phi\right)$, is related to the earlier considered z-axis traveling plane wave, $h_{ab},$ by $h'_{ab}\left(\theta,\phi\right) = \left(R^{-1}\right)^i{}_{a}h_{ij}R^j{}_{b}.$ The response, $f$, of a LIGO-like interferometer is then
        \begin{align}
        	f\left(\theta,\phi\right) &= \tfrac{1}{2}h'_{ab}\left(\theta,\phi\right)A^{ab},\text{where}\\
        	A &= \begin{bmatrix}
        		0 & 0 & 0 & 0\\
        		0 & l^x & 0 & 0\\
        		0 & 0 & -l^y & 0\\
        		0 & 0 & 0 & 0
        	\end{bmatrix},\text{in (t,x,y,z) order,}
        \end{align}
        where $l^x$ and $l^y$ correspond to the length of the respective arms of the interferometer situated on the axis of the x-y plane. For convenience, parameters were chosen to emphasize the features of the radiation patterns and $l^x = l^y.$ The radiation pattern for each species is presented.
        
        \subsubsection{Scalar Antenna Pattern}
        
        \begin{align}
                f_{\text{scalar}}\left(\theta,\phi\right) &\propto \cos\left(2\phi\right)\sin^2\theta
        \end{align}
        
        \begin{figure}[h!]
                \caption{Example of a LIGO-like gravitational Michelson interferometer like detector's directional sensitivity to a scalar diffeomorphism fluctuation in the z-direction.}
                \centering
                \includegraphics[width=0.5\textwidth]{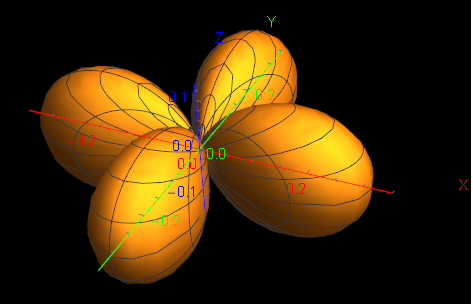}
        \end{figure}
        
        \subsubsection{Solenoidal Antenna Pattern}
        
        \begin{align}
                f_{\text{Sol1}}\left(\theta,\phi\right) \propto& 4\tfrac{k\lambda_0^2}{\kappa_0}\cos\left(\theta-2\phi\right) + 15J_0k^2\lambda_0^2\cos\left(2\left(\theta-\phi\right)\right)\cr
                & + \left(1200J_0 - 30J_0k^2\lambda_0^2\right)\cos\left(2\phi\right)\cr
                & + 15J_0k^2\lambda_0^2\cos\left(2\left(\theta+\phi\right)\right) - 4\tfrac{k\lambda_0^2}{\kappa_0}\cos\left(\theta+2\phi\right)
        \end{align}
        
        \begin{figure}[h!]
                \caption{Example of a LIGO-like gravitational Michelson interferometer like detector's directional sensitivity to a solenoidal diffeomorphism fluctuation in the z-direction from the $P_1$ solenoidal degree of freedom.}
                \centering
                \includegraphics[width=0.5\textwidth]{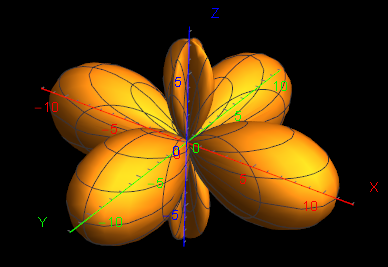}
        \end{figure}
        
        \begin{align}
                f_{\text{Sol2}}\left(\theta,\phi\right) \propto& \Big(15J_0\kappa_0\left(k^2\lambda_0^2-20\right)\cos2\theta + 2k\lambda_0^2\sin2\theta\cr
                & - 15J_0\kappa_0\left(k^2\lambda_0^2+20\right)\Big)\cos2\phi
        \end{align}
        
        \begin{figure}[h!]
                \caption{Example of a LIGO-like gravitational Michelson interferometer like detector's directional sensitivity to a solenoidal diffeomorphism fluctuation in the z-direction from the $P_2$ solenoidal degree of freedom.}
                \centering
                \includegraphics[width=0.5\textwidth]{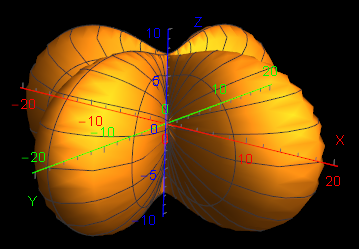}
        \end{figure}
        
        \begin{align}
                f_{\text{Sol3}}\left(\theta,\phi\right) \propto&  \cos\left(2\phi\right)\sin^2\theta
        \end{align}
        
        \begin{figure}[h!]
                \caption{Example of a LIGO-like gravitational Michelson interferometer like detector's directional sensitivity to a solenoidal diffeomorphism fluctuation in the z-direction from the $P_3$ solenoidal degree of freedom.}
                \centering
                \includegraphics[width=0.5\textwidth]{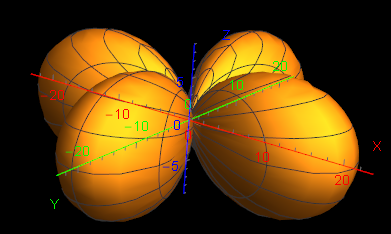}
        \end{figure}
        
        \subsubsection{Transverse Antenna Pattern}
        
        \begin{align}
                &f_{\hat{\mathcal{P}}_{xx}}\left(\theta,\phi\right) \propto \Big(9-2J_0\kappa_0\left(k^2\lambda_0^2-30\right)\cr
                & + \left(3+jJ_0\kappa_0\left(k^2\lambda_0^2-30\right)\right)\cos2\theta\Big)\cos2\phi
        \end{align}
        
        \begin{figure}[h!]
                \caption{Example of a LIGO-like gravitational Michelson interferometer like detector's directional sensitivity to a transverse diffeomorphism fluctuation in the z-direction from the $\left(\mathcal{P}_{TT}\right)_{xx}$ transverse degree of freedom.}
                \centering
                \includegraphics[width=0.5\textwidth]{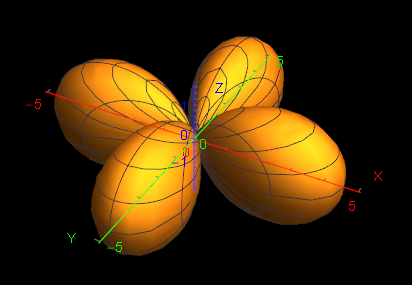}
        \end{figure}
        
        \begin{align}
                f_{\hat{\mathcal{P}}_{xy}}\left(\theta,\phi\right) \propto& J_0\kappa_0\left(k^2\lambda_0^2-30\right)\cos2\left(\theta-\phi\right)\cr
                &-2J_0\kappa_0\left(k^2\lambda_0^2-30\right)\cos2\phi\cr
                &+J_0\kappa_0\left(k^2\lambda_0^2-30\right)\cos2\left(\theta+\phi\right)\cr
                &+6\left(\sin\left(\theta-2\phi\right) - \sin\left(\theta+2\phi\right)\right)
        \end{align}
        
        \begin{figure}[h!]
                \caption{Example of a LIGO-like gravitational Michelson interferometer like detector's directional sensitivity to a transverse diffeomorphism fluctuation in the z-direction from the $\left(\mathcal{P}_{TT}\right)_{xy}$ transverse degree of freedom.}
                \centering
                \includegraphics[width=0.5\textwidth]{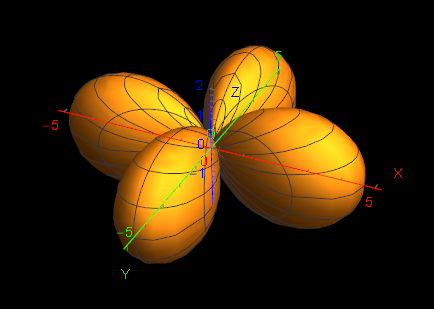}
        \end{figure}
        
        \begin{align}
                &f_{\hat{\mathcal{P}}_{xz}}\left(\theta,\phi\right) \propto 5J_0\kappa_0\Big(-30-k^2\lambda_0^2\cr
                & +\left(k^2\lambda_0^2-30\right)\cos2\theta\Big)\left(\cos^2\theta - \sin^2\theta\right)\cr
                & + 2\left(k^2\lambda_0^2 -15\right)\sin\theta\sin2\phi
        \end{align}
        
        \begin{figure}[h!]
                \caption{Example of a LIGO-like gravitational Michelson interferometer like detector's directional sensitivity to a transverse diffeomorphism fluctuation in the z-direction from the $\left(\mathcal{P}_{TT}\right)_{xz}$ transverse degree of freedom.}
                \centering
                \includegraphics[width=0.5\textwidth]{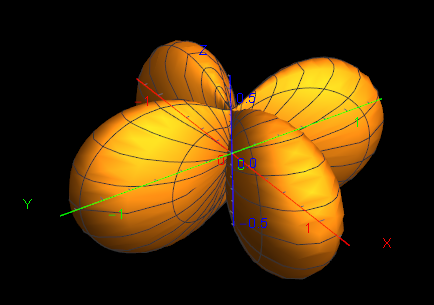}
        \end{figure}
        
        \begin{align}
                f_{\hat{\mathcal{P}}_{yz}}\left(\theta,\phi\right) \propto& \Big(-5J_0\kappa_0\left(k^2\lambda_0^2-60\right) + 5J_0\kappa_0\lambda_0^2\cos2\theta\cr
                & + \left(k^2\lambda_0^2-15\right)\sin2\theta\Big)\cos2\phi
        \end{align}
        
        \begin{figure}[h!]
                \caption{Example of a LIGO-like gravitational Michelson interferometer like detector's directional sensitivity to a transverse diffeomorphism fluctuation in the z-direction from the $\left(\mathcal{P}_{TT}\right)_{yz}$ transverse degree of freedom.}
                \centering
                \includegraphics[width=0.5\textwidth]{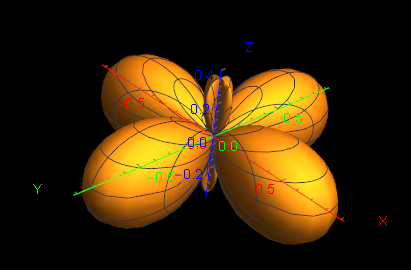}
        \end{figure}
        
        \begin{align}
                &f_{\hat{\mathcal{P}}_{zt}}\left(\theta,\phi\right) \propto \Big(-150J_0\kappa_0\left(k^2\lambda_0^2-60\right)\sqrt{k^2-\tfrac{15}{\lambda_0^2}}\cr
                & + k^3\lambda_0^2\left(k^2\lambda_0^2-15\right)\cr
                & - k\big(450-45k^2\lambda_0^2 - 150J_0k\kappa_0\lambda_0^2\sqrt{k^2-\tfrac{15}{\lambda_0^2}}\cr
                & + k^4\lambda_0^4\big)\cos2\theta\Big)\cos2\phi
        \end{align}
        
        \begin{figure}[h!]
                \caption{Example of a LIGO-like gravitational Michelson interferometer like detector's directional sensitivity to a transverse diffeomorphism fluctuation in the z-direction from the $\left(\mathcal{P}_{TT}\right)_{zt}$ transverse degree of freedom.}
                \centering
                \includegraphics[width=0.5\textwidth]{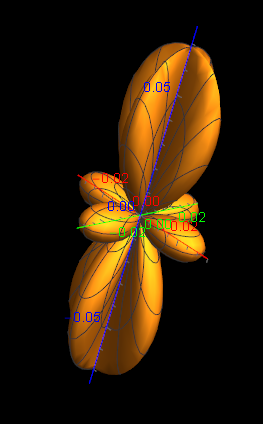}
        \end{figure}
        
        \section{Conclusion and Future Work}
        
        TW gravity has degrees of freedom from the metric, the diffeomorphism field, and the projective invariant. The diffeomorphism field itself can be further decomposed in transverse-traceless, solenoidal-traceless, and trace degrees of freedom. The transverse-traceless and solenoidal-traceless fields are tachyonic as free fields. We have considered geodesic deviation sourced by plane-waves in each of these sectors: firstly in a Minkowski space background, and later with back-reacted contributions to linear order in $h_{ab}$ sourced by each diffeomorphism fields radiation species. The diffeomorphism field radiation produces trivial geodesic deviation in Minkowski space background, but produces a longitudinal contribution to deviation when back reaction to metric degrees of freedom are considered to linear order.
An antennae design such a LISA could detect the presence of these scalar modes as well as longitudinal modes.  This work demonstrates how  the diffeomorphism field can enter into gravitational fluctuation and produce gravitational radiation outcomes.  One then asks, can the diffeomorphism field mask the signature of black hole and neutron star mergers and can it provide primordial seeds to gravitational density fluctuations in the early universe. \         
        \section*{Acknowledgment}
        
        The authors would like to thank the Diffeomorphisms and Geometry Research Group, the
        Nuclear and Particle Theory Group at the University of Iowa, and especially Dr. Shammi Tahura for constructive conversations throughout this work.
        
        \appendix
        
        \section{Definitions}
        
        We use a metric signature of mostly minus (+,-,-,-). Under case Latin indices $a,b,\dots$ range over the space-time coordinates $0,1,\dots,d-1$ where as upper case Latin indices $A,B,\dots$ range over the Thomas cone coordinates $0,1,\dots,d-1,\lambda$. We use parentheses and brackets to denote symmetrization and anti
symmetrization:
        
        \begin{align}
                T_{(ab)} = T_{ab} + T_{ba},~~ T_{[ab]} = T_{ab} - T_{ba}.
        \end{align}
        
        The connection symbol  $\Gamma^a{}_{bc}$ on the spacetime manifold is independent of the metric a la Palatini, while $\hat{\Gamma}^a{}_{bc}$ is the usual Levi-Civita related connection. They differ by the Palatini field $C^a{}_{bc}:$
        
        \begin{align}
                \Gamma^a{}_{bc} = \hat{\Gamma}^a{}_{bc} + C^a{}_{bc}.
       \label{Cdefinition}  \end{align}
        
        \noindent The connection components of the Thomas cone are given by
        
        \begin{equation}
                \label{eq: Gammatilde}
                \tilde{\Gamma}^{A}{}_{BC} = \begin{cases}
                        \tilde{\Gamma}^a{}_{bc} = \Pi^a{}_{bc} \\
                        \tilde{\Gamma}^\lambda{}_{bc} = \Upsilon^\lambda \mathcal{D}_{bc} \\ 
                        \tilde{\Gamma}^a{}_{b\lambda} = \tilde{\Gamma}^a{}_{\lambda b} = \alpha_\lambda \delta^a{}_b \\ 
                        \tilde{\Gamma}^\lambda{}_{b\lambda} = \tilde{\Gamma}^a{}_{\lambda b} = \tilde{\Gamma}^\lambda{}_{\lambda\lambda} = 0 \\ 
                \end{cases}\, 
        \end{equation}
        where 
        \begin{align}
                \label{e:Pi}
                {\Pi}^{a}_{\,\,\,\,b c} =& { \G}^{a}_{\,\,\,\,b c} + \delta^a{}_{(c}~ \a_{b)}~~~,~~~\a_a = -\tfrac{1}{\rd +1} \Gamma^m{}_{am}\\
                \Upsilon^A =& (0,0,\dots,0,\lambda)~~~,~~~\a_\b = \left( \a_c, \lambda^{-1} \right). \label{UpsilonOmega}
        \end{align}
        
        A coordinate transformation on the spacetime and its Jacobian are given by
        
        \begin{align}
                x'^m = x'^m(x^n),~~ J^m{}_{n} = \tfrac{\partial x'^m}{\partial x^n} \label{TCTransformations}
        \end{align}
while the Thomas cone volume coordinate transforms as $\lambda' = \lambda |J|^{-\tfrac{1}{d+1}}$ where $J$ is the determinant of the Jacobian of the transformation.  The volume form $g_{a}$ being the trace of the Levi-Civita connection: 
\begin{align}
                g_{a} = -\tfrac{1}{d+1}\tilde{\Gamma}^m{}_{bm} = -\tfrac{1}{d+1}\partial_{a}\log\sqrt{|g|}.
        \end{align}
        
        \noindent The soldering metric for vectors on the Thomas cone and its inverse are given by
        
        \begin{align}
                G_{AB} = & 
                \begin{bmatrix}
                        g_{ab}-\lambda_0^{\ 2} g_a g_b & -\frac{\lambda_0^{\ 2}}{\lambda} g_a \\
                        -\frac{\lambda_0^{\ 2}}{\lambda} g_b & -\frac{\lambda_0^{\ 2}}{\lambda^2}
                \end{bmatrix} \\ 
                G^{AB} = & 
                \begin{bmatrix}
                        g^{ab} & -\lambda g^{am} g_m \\
                        -\lambda g^{bm}g_m & \frac{\lambda^2}{\lambda_0^{\ 2}}\left(-1 + g^{mn}\lambda_0^{\ 2}g_m g_n \right)
                \end{bmatrix}. 
        \end{align}
        
        Through the use of $\Upsilon,$ we may write this as
        
        \begin{align}
                G_{AB} &= \delta^a_{\,\,\alpha} \delta^b_{\,\,\beta} \,g_{ab} - \lambda_0^2 g_\alpha g_\beta \\
                G^{AB} &= g^{ab} (\delta^\alpha_{\ a} - g_a \Upsilon^\alpha)(\delta^\beta_{\ b} - g_b \Upsilon^\alpha) - \lambda_0^{-2} \Upsilon^A \Upsilon^B. 
        \end{align}
        
        \noindent To discuss projectively invariant tensors, we introduce $\Vin_a = g_a - \a_a = \tfrac{1}{d+1}C^m{}_{am}.$ $\tilde{\nabla}_A$ is the covariant derivative on the Thomas cone related to $\Gamma^A{}_{bc}.$ The covariant derivatives $\nabla_a$ and $\hat{\nabla}_a$ use $\Gamma^a{}_{bc}$ and $\hat{\Gamma}^a{}_{bc}$ respectively. The derivative $\breve{\nabla}$ while not covariant is defined similarly with $\Pi^a{}_{bc}$ used as a connection. The covariant derivative $\bar{\nabla}_a$ maintains projective invariance of derived tensors. A summary of connection symbols is given by
        
        \begin{align}
                \tilde{\Gamma}^a{}_{bc} &= \tfrac{1}{2}g^{al}\left[\partial_{b}g_{lc} + \partial_{c}g_{bl} - \partial_lg_{bc}\right]\\
                \Gamma^a{}_{bc} &= \tilde{\Gamma} + C^{a}{}_{bc}\\
                \tilde{C}^a{}_{bc} &= C^a{}_{bc} - \tfrac{1}{d+1}\delta^a{}_{(b}C^{m}{}_{c)m} = C^a{}_{bc} - \delta^a{}_{(b}\Vin_{c)}\\
                \Pi^a{}_{bc} &= \hat{\Gamma}^a{}_{bc} + \delta^a{}_{(b} g_{c)} + \tilde{C}^a{}_{bc}\\
                \bar{\Gamma}^a{}_{bc} &= \hat{\Gamma}^a{}_{bc} + \tilde{C}^a{}_{bc} = \Gamma^a{}_{bc} - \delta^a{}_{(b}\Vin_{c)}
        \end{align}
        
        \noindent with associated curvature tensors defined as
        
        \begin{align}
                \hat{R}^{a}{}_{bmn} = \hat{\Gamma}^{a}{}_{b[n,m]} &+ \hat{\Gamma}^{c}{}_{b[n}\hat{\Gamma}^{a}{}_{m]c}\\
                Q^{a}{}_{bmn} = C^{a}{}_{b[n,m]} &+ C^{c}{}_{b[n}C^{a}{}_{m]c}\\
                R^{a}{}_{bmn} = \Gamma^{a}{}_{b[n,m]} &+ \Gamma^{c}{}_{b[n}\Gamma^{a}{}_{m]c}\\
                \mathcal{R}^{a}{}_{bmn} = \Pi^{a}{}_{b[n,m]} &+ \Pi^{c}{}_{b[n}\Pi^{a}{}_{m]c}\\
                \bar{R}^{a}{}_{bmn} = \bar{\Gamma}^{a}{}_{b[n,m]} &+ \bar{\Gamma}^{c}{}_{b[n}\bar{\Gamma}^{a}{}_{m]c}.
        \end{align}
        
        By contracting the first and third indices, and further contracting indices with the metric the following quantities are defined:
\begin{align}
                \hat{R}_{bd} = \hat{R}^a{}_{bad},~R_{bd} = R^a{}_{bad},~Q_{bd} = Q^a{}_{bad},~\cr
                \mathcal{R}_{bd} = \mathcal{R}^a{}_{bad},~\bar{R}_{bd} = \bar{R}^a{}_{bad}\\
                \hat{R} = \hat{R}_{bd}g^{bd},~R = R_{bd}g^{bd},~Q = Q_{bd}g^{bd},~\cr
                \mathcal{R} = \mathcal{R}_{bd}g^{bd},~\bar{R} = \bar{R}_{bd}g^{bd}.
        \end{align}
Now, while the Diffeomorphism field, $\mathcal{D}_{ab}$, is not a tensor as it is a component of the Thomas cone connection, it's degrees of freedom can be cast into a tensor $\mathcal{P}_{ab}$,
\begin{align}
                \mathcal{P}_{ab} = \mathcal{D}_{ab} - \partial_{a}\alpha_b + \Gamma^c{}_{ab}\alpha_c + \alpha_a\alpha_b.
   \label{Pdefinition}     \end{align}
 However, $\mathcal{P}_{ab}$ is not invariant under projective transformation. For a projectively invariant tensor, $\bar{\mathcal{P}}_{ab}$ is defined.
\begin{align}
                \bar{\mathcal{P}}_{ab} &= \mathcal{P}_{ab} - \nabla_a\Vin_b - \Vin_a\Vin_b\cr
                & = \mathcal{D}_{ab} - \breve{\nabla}_{a}g_{b} -g_ag_b
        \end{align}
The projective curvature tensor for the Thomas cone is constructed from the projective connection $\tilde{\Gamma}$ and it's associated covariant derivative $\tilde{\nabla}$;
 \begin{align}
                \left[\tilde{\nabla}_{A},\tilde{\nabla}_{B}\right]V^{\gamma} &= \mathcal{K}^{C}{}_{DAB}V^D\\
                \left[\tilde{\nabla}_{A},\tilde{\nabla}_{B}\right]V_{\gamma} &= -\mathcal{K}^{D}{}_{CAB}V_D\\
                \mathcal{K}^{A}{}_{BMN} = \tilde{\Gamma}^{A}{}_{B[N,M]} &+ \tilde{\Gamma}^{C}{}_{B[N}\tilde{\Gamma}^{A}{}_{M]C}.   
        \end{align}
The nonvanishing curvature tensors can be written in terms of equivalent formulations by
\begin{align}
                \mathcal{K}^a{}_{bcd} &= \mathcal{R}^a{}_{bcd} + \delta^a_{[c}\mathcal{D}_{d]b}\cr
                &= R^a{}_{bcd} + \delta^a_{[c}\mathcal{P}_{d]b} - \delta^a{}_b\mathcal{P}_{[cd]}\cr
                &= \hat{R}^a{}_{bcd} + Q^{a}{}_{bcd} + \delta^a_{[c}\mathcal{P}_{d]b} - \delta^a{}_b\mathcal{P}_{[cd]}\cr
                &= \bar{R}^a{}_{bcd} + \delta^a_{[c}\bar{\mathcal{P}}_{d]b}
        \end{align}
 \begin{align}
                \mathcal{K}_{bd} &= \mathcal{R}_{bd} + \left(d-1\right)\mathcal{D}_{bd}\cr
                &= R_{bd} + \left(d-1\right)\mathcal{P}_{bd} - d\mathcal{P}_{[bd]}\cr
                &= \hat{R}_{bd} + Q_{bd} + \left(d-1\right)\mathcal{P}_{bd} - d\mathcal{P}_{[bd]}\cr
                &= \bar{R}_{bd} + \left(d-1\right)\bar{\mathcal{P}}_{bd}
        \end{align}
\begin{align}
                \mathcal{K} &= \mathcal{R} + \left(d-1\right)\mathcal{D}\cr
                &= R + \left(d-1\right)\mathcal{P}\cr
                &= \hat{R} + Q + \left(d-1\right)\mathcal{P}\cr
                &= \bar{R} + \left(d-1\right)\bar{\mathcal{P}}
        \end{align}
        
        \begin{align}
                K_{bcd} &= \breve{\nabla}_{[c}\mathcal{D}_{d]b} + g_a\mathcal{K}^a{}_{bcd}\cr
                &= \nabla_{[c}\mathcal{P}_{d]b} + \Vin_a\mathcal{K}^a{}_{bcd}\cr
                &= \bar{\nabla}_{[c}\bar{\mathcal{P}}_{d]b}
        \end{align}
        
%
\nocite{*}
\bibliographystyle{apsrev4-1} 
\bibliography{References} 
\end{document}